\documentclass{iopart}             
\usepackage{iopams}                
\usepackage{epsf,epsfig,graphics}
\input{psfig}
\DeclareMathAlphabet{\scriptnew}{U}{eus}{m}{n}
\def\rma{{\rm a}}
\def\rmb{{\rm b}}
\def\rmc{{\rm c}}
\def\rme{{\rm e}}
\def\rms{{\rm s}}
\def\rmp{{\rm p}}
\def\rmd{{\rm d}}
\def\rmf{{\rm f}}
\def\rmg{{\rm g}}

\def\rmS{{\rm S}}
\def\rmP{{\rm P}}
\def\rmD{{\rm D}}
\def\rmF{{\rm F}}
\def\rmH{{\rm H}}

\def\rmL{{\rm L}}

\def\AA{{\it Astron. Astrophys. }}
\def\AASS{{\it Astron. Astrophys. Suppl. Ser. }}
\def\APJ{{\it Astrophys. J. }}
\def\APJS{{\it Astrophys. J. Suppl. Ser. }}
\begin{document}
\setlength{\arraycolsep}{2.5pt}             
\jl{2}
\setcounter{footnote}{2}
\title[Dielectronic recombination of Fe$^{13+}$]
       {Dielectronic recombination of Fe$^{13+}$: benchmarking the M-shell }
\author{N R Badnell}
\address{Department of Physics, University of Strathclyde, Glasgow, G4 0NG, UK}

\begin{abstract}
We have carried-out a series of multi-configuration Breit--Pauli {\sc autostructure}
calculations for the dielectronic recombination of Fe$^{13+}$. We present a detailed comparison
of the results with the high-energy resolution measurements reported recently from the
Heidelberg Test Storage Ring by Schmidt \etal$\!$. Many Rydberg series contribute
significantly from this initial $3\rms^2 3\rmp$ M-shell ion, resulting in a complex
recombination `spectrum'. While there is much close agreement between theory and experiment, 
differences of typically 50\% in the summed resonance strengths over $0.1-10$~eV result in the
experimentally based total Maxwellian recombination rate coefficient being a factor
of $1.52-1.38$ larger than theory over $10^4-10^5$~K, which is a typical temperature range of
peak abundance for Fe$^{13+}$ in a photoionized plasma. Nevertheless, this theoretical recombination
rate coefficient is an order of magnitude larger than that used by modellers to-date.
This may help explain the discrepancy between the iron M-shell ionization balance 
predicted by photoionization modelling codes such as {\sc ion} and {\sc cloudy}
and that 
deduced from the iron M-shell
unresolved-transition-array absorption feature observed in the X-ray spectrum of many active galactic nuclei.
Similar data are required for Fe$^{8+}$ through Fe$^{12+}$ to remove the question mark hanging
over the atomic data though.
\end{abstract}
\submitted
\section{Introduction}
Dielectronic recombination (Burgess 1964) is the dominant electron--ion recombination process in
both photoionized and collisional plasmas. Extensive theoretical data is available
for all K- and L-shell ions of all elements up to Zn, and selected heavy elements
beyond, following the work of Badnell \etal (2003), and is available online\footnote{Webpage
http://amdpp.phys.strath.ac.uk/tamoc/DATA}. These data, including radiative recombination (RR),
have been used to provide new ionization balances for both electron collisional (Bryans \etal 2006)
and photoionized plasmas (Ferland 2006). Extensive benchmarking has taken place against experiment for both 
low-$Z$ (C, N, O) and higher-$Z$ elements (Fe, Ni) --- see e.g. Fogle \etal (2005) and Savin \etal (2006), 
and references therein. Work on the M-shell is sparse (beyond the simple Na-like sequence, 
Linkemann \etal 1995, Fogle \etal 2003). Yet, M-shell Fe ions are ubiquitous in astrophysics. 
It has become clear recently that dielectronic recombination (DR) rate coefficients for 
Fe $3\rmp^q (q=1-6)$ ions (Fe$^{8+}$ -- Fe$^{12+}$) are highly questionable at temperatures where these
ions form in photoionized plasmas ($10^4-10^5$~K, say, Kallman and Bautista 2001).
This stems from the inability of photoionized plasma modelling codes such as {\sc ion} (Netzer 2004)
and {\sc cloudy} (Kraemer \etal 2004) to model the iron M-shell unresolved-transition-array 
absorption feature observed in the X-ray spectrum of many active galactic nuclei. The situation can be improved 
by changing the ionization balance for these Fe ions at such temperatures, as first suggested by
Netzer \etal (2003), and which is achieved by increasing the dielectronic recombination rate coefficients by
large factors (e.g., 2 -- 4). That this is a plausible approach has been verified experimentally 
for Fe$^{13+}$ by Schmidt \etal (2006) who carried-out high-energy resolution DR measurements at the
Heidelberg Test Storage Ring (TSR). They deduced a Maxwellian recombination rate coefficient which is 
up to an order of magnitude larger than that recommended by Arnaud \& Raymond (1992) and Mazzotta \etal (1998)
at photoionized plasma temperatures, and which is currently used to determine the ionization balance 
of iron in {\sc ion} and {\sc cloudy}, and other photoionized plasma modelling codes such as {\sc xstar} 
(Kallman and Bautista 2001). 
The reason for this difference is that the existing theoretical dielectronic recombination contribution
(Jacobs \etal 1977) to the total recombination rate coefficient falls-off exponentially below 
$\sim 10$~eV and the total is dominated by direct radiative recombination. A similar problem was noted by
M\"{u}ller (1999) for Fe$^{15+}$ and it is prevelant also for L-shell ions, following the pioneering work
of Savin \etal (1997). The approach of Jacobs 
\etal (1977) is based upon the `no-coupling' scheme, allows only for dipole core-excitations in 
the dielectronic capture process and pays no detailed attention to the positioning of near-threshold 
resonances. It should be noted that the work of Jacobs \etal (1977) was motivated by applications to
high-temperature electron collision dominated plasmas and, for such, their approach is quite reasonable.
It is clear, however, that the DR of Fe$^{13+}$ needs to be re-examined for application
to photoionized plasmas. Such a re-examination, including a comparison with the results
of the measurements by Schmidt \etal (2006), will provide a benchmark for other 
Fe $3\rmp^q (q=1-6)$ ions, and the M-shell more generally.

The outline of the rest of this paper is as follows: in section 2 we describe our theoretical
approach; in section 3 we make a detailed study of the structure of Fe$^{13+}$; in section 4 we  
compare our velocity-convoluted DR cross sections with those from the experiment by Schmidt 
\etal (2006); and in section 5 we compare various Maxwellian recombination rate coefficients.

\section{Theory}
\label{thy}
We have used {\sc autostructure} (Badnell 1987, 1997) to carry-out a series of multi-configuration
Breit--Pauli calculations of dielectronic recombination cross sections and rate coefficients.
The method implemented within {\sc autostructure} is the independent processes, isolated resonances 
using distorted waves (IPIRDW) approach to DR.  A detailed discussion of the validity of
this approach is given by Pindzola \etal 1992) while its advantages from a (collisional--radiative) 
modelling perspective is discussed by Badnell \etal (2003). 

Let $\sigma^{j}_{f\nu}(E)$ denote the partial dielectronic recombination cross section, as a
function of center-of-mass energy $E$,
from an initial metastable state $\nu$ of an ion $X^{+z}$, through an autoionizing state $j$,
into a resolved final state $f$ of an ion $X^{+z-1}$, then
\begin{eqnarray}
\label{Lor}
\sigma^{j}_{f\nu}(E)&=&\hat{\sigma}^{j}_{f\nu}L^j(E),
\end{eqnarray}
where $L^j(E)$ is the Lorentzian line shape of the resonance (energy-normalized to unity). 
Here, $\hat{\sigma}$
denotes the integrated (partial) dielectronic recombination cross section, which is given by
\begin{eqnarray}
\hat{\sigma}^{j}_{f\nu}(E_\rmc)&=&{\left(2\pi a_0 I_{\rm H}\right)^{2} \over E_\rmc }
{\omega_{j} 
\over 2\omega_{\nu}}
{ \tau_0\sum_{l}A^{{\rm a}}_{j \rightarrow \nu, E_{\rmc}l} \, A^{{\rm r}}_{j \rightarrow f}
\over \sum_{h} A^{{\rm r}}_{j \rightarrow h} + \sum_{m,l} A^{{\rm a}}_{j \rightarrow m, E_{\rmc}l}}\, ,
\label{pdri}
\end{eqnarray}
where $\omega_j$ is the statistical weight of the
$(N+1)$-electron doubly-excited resonance state $j$, $\omega_\nu$ is the statistical weight
of the $N$-electron target state (so, $z=Z-N$, where $Z$ is the nuclear charge) and the autoionization 
($A^{\rm a}$) and radiative
($A^{\rm r}$) rates are in inverse seconds. Here, $E_{\rmc}$ is the energy of the continuum 
electron (with orbital angular momentum $l$), which is fixed by the position of the resonance $j$
relative to the continuum $\nu$, 
$I_{\rm H}$ is the ionization potential energy of the hydrogen atom (both in the
same units of energy)
and $(2\pi a_0)^{2}\tau_0=2.6741\times10^{-32}$ cm$^2$s. 

A powerful aspect of the IPIRDW approach is that the use of equation (\ref{Lor}) enables
an analytic integration over the resonance profiles to be carried-out. This is in contrast
to an $R$-matrix calculation which must map-out the detailed resonance structure numerically.
This in itself is more demanding for DR than for electron-impact excitation since a much finer 
energy mesh is needed to map-out all resonances which contribute significantly to the cross section
--- see Gorczyca \etal (2002) for a detailed study and discussion of the issue. 

So, let
$\bar{\sigma}^{j}_{f\nu}$ denote the corresponding energy-averaged (partial) dielectronic 
recombination cross section, then
\begin{eqnarray}
\bar{\sigma}^{j}_{f\nu}(E_\rmc)&\equiv&{1\over \Delta E} \int^{E_\rmc+\Delta E/2}_{E_\rmc-\Delta E/2} 
{\sigma}^{j}_{f\nu} (E')\rmd E'\,.
\end{eqnarray}
Here, $\Delta E$ denotes the bin width energy, which is chosen so as to be large compared to
the Lorentzian width and small compared to the characteristic width of any subsequent convolution; 
otherwise, the choice of $\Delta E$ is arbitrary and it is usually taken to be a constant. Then,
\begin{eqnarray}
\label{bine}
\bar{\sigma}^{j}_{f\nu}(E_\rmc)&=&{1\over \Delta E} \hat{\sigma}^{j}_{f\nu}(E_\rmc)\,.
\end{eqnarray}
Thus, for a fixed $j$ and  $\nu$, the energy-averaged partial DR cross section takes-on a non-zero
value at a single energy, $E_\rmc$, including when summed-over final states $f$. Most applications 
involve a sum over resonance levels $j$ and it is convenient to `bin' the cross section via
\begin{eqnarray}
\bar{\sigma}_{\nu}(E_m)&=&\sum_{f,j} \bar{\sigma}^{j}_{f\nu}(E_\rmc)\quad\quad \forall\quad E_\rmc 
\in [E_m, E_{m+1})\,,
\end{eqnarray}
where $E_{m+1}=E_m+\Delta E$. The sum over $f$ is over all final states which lie below the
ionization limit and which may include cascade through autoionizing levels, although a
single cascade (i.e. a two-step radiative stabilization) is usually more than sufficient.
For total rate coefficients, applicable low-density plasmas, the sums over $f$ and $j$
are taken to convergence but for application to laboratory measurements the sum over $f$
(and hence, in practice, $j$) is truncated.

\subsection{Application to merged-beams measurements}
Merged-beams measurements utilizing an electron-cooler determine a rate coefficient for the
dielectronic recombination process. To compare with a measurement at an electron--ion centre-of-mass energy 
$E_0$, we determine a corresponding theoretical rate coefficient, $\alpha(v_0)$, formally given by
\begin{eqnarray}
\label{exp1}
\alpha(v_0)&=&<v\sigma>=\int \sigma(v)vf(v_0,{\bf v})\rmd{\bf v}\,,
\end{eqnarray}
where $f(v_0,{\bf v})$ is the merged-beams electron velocity distribution in the center-of-mass frame
of the ions and $v_0= \sqrt{2E_0/m_\rme}$, since the electrons are moving 
non-relativistically with mass $m_\rme << m_X$, the mass of the ion $X$.

The experimental velocity distribution, $f(v_0,{\bf v})$, is a `flattened Maxwellian' (Dittner \etal 1986) which is 
characterized  by two parameters, a `parallel' temperature $T_\parallel$ and `perpendicular' 
temperature $T_\perp$, with $T_\parallel << T_\perp$:
\begin{eqnarray}
\label{exp2}
\fl
f(v_0,{\bf v})&=&\left({m_\rme \over 2\pi kT_\parallel}\right)^{1/2}\exp\left[-{m_\rme\left(v_\parallel-v_0\right)^2
\over 2kT_\parallel}\right]
{m_\rme \over 2\pi kT_\perp}\exp\left(-{m_\rme v^2_\perp \over 2kT_\perp}\right)\,,
\end{eqnarray}
where $v_\parallel$ and $v_\perp$ denote the parallel and perpendicular components of ${\bf v}$, respectively.
Note, at high energies, $E_0>>kT_\perp$, $f(v_0,{\bf v})$ reduces to an effective  Gaussian distribution with a
full-width at half-maximum of $2(\ln 2\,E_0 k T_\parallel)^{1/2}$.

For a bin width that is much smaller than the energy resolution of the experiment,
and on using the distribution given by equation (\ref{exp2}), we can
write equation (\ref{exp1})  in terms of the 
energy-averaged cross sections and bin energies, $E_m=m_\rme v_m^2/2$:
\begin{eqnarray}
\label{exp3}
\fl
\alpha_\nu(v_0)=\sum_m{\Delta E \bar{\sigma}_{\nu}(E_m)v_m \over 2kT_\perp\sqrt{\left(1-T_\parallel/T_\perp\right)}}
\exp\left({E_0 \over \left(kT_\perp - kT_\parallel\right)}-{E_m \over kT_\perp}\right)\\ \nonumber
\lo
\times \left[ {\rm erf}\left(z_1-z_2\right)+{\rm erf}\left(z_1+z_2\right)\right]\,,
\end{eqnarray}
where
\begin{eqnarray}
z_1=\left[{E_m\left(kT_\perp-kT_\parallel\right) \over kT_\perp kT_\parallel}\right]^{1/2}
\end{eqnarray}
and
\begin{eqnarray}
z_2=\left[{E_0 kT_\perp \over kT_\parallel\left(kT_\perp-kT_\parallel\right)}\right]^{1/2}\,.
\end{eqnarray}
Recall, $\bar{\sigma}_{\nu}$ has units of cm$^2$. Writing $v_m/2=\sqrt{E_m/I_\rmH}\sqrt{I_\rmH/(2m_\rme)}$,
we have that $\sqrt{I_\rmH/(2m_\rme)}=1.0938\times 10^8$ cm s$^{-1}$ is the relevant remaining constant
which defines the rate coefficient.

In measurements carried-out at storage rings, the ions are circulated long enough,
and the densities are low enough, for the ion population to be concentrated in
the ground state, normally. In single-pass measurements it is necessary to calculate DR 
cross sections for metastable levels as well and then to combine them using experimental
metastable fractions, if possible, or, typically, to use fractions for which the
resulting cross section best matches the measured.

\subsubsection{Survival of the species.}
Recombined ions with high principal quantum numbers are re-ionized by the strong
electric field present in the charge-state analyzer which is used to separate the recombined ions 
from the original ion beam, and so are not counted as recombined ions. A `hard' cut-off at $n_\rmc$,
given by the hydrogenic expression (Bethe and Salpeter 1957)
\begin{eqnarray}
\label{ncut}
n_\rmc=\left(6.2\times 10^8 z^3/F\right)^{1/4}\,,
\end{eqnarray}
where $F$ (V/cm) is the field strength, often suffices. Sometimes, however, recombined ions
with $n>n_\rmc$ have time to radiatively stabilize to $n<n_\rmc$ during the time-of-flight ($\tau_\rmF$)
from the cooler to the analyzer and so survive to be counted. This is modelled theoretically through 
the use of a `soft' or `delayed' cut-off  (Zong \etal 1998, Schippers \etal 2001). A soft cut-off 
simply imposes a higher effective $n_\rmc$ based-upon the lifetime of the Rydberg states. The delayed
cut-off approach determines, for $n>n_\rmc$, the lifetime of each Rydberg state (usually hydrogenic),
$\tau_{nl}$, and multiplies each $nl$ partial DR cross section by a survival probability, given by
\begin{eqnarray}
P_{nl}=1-{\tau_{nl}\over \tau_\rmL}\exp\left({-\tau_\rmF\over\tau_{nl}}\right)\left[\exp\left({\tau_\rmL\over 
2\tau_{nl}}\right)-\exp\left({-\tau_\rmL\over 2\tau_{nl}}\right)\right]\,,
\end{eqnarray}
where $\tau_\rmL$ is the time-of-flight for the passage through the merged-beams section of the
cooler. It is often the case that $\tau_\rmL\ll \tau_{nl}$ for the contributing $nl$ and so,
to a good approximation,
\begin{eqnarray}
\label{prob}
P_{nl}=1-\exp\left({-\tau_\rmF\over\tau_{nl}}\right)\,,
\end{eqnarray}
i.e., independent of the cooler time-of-flight.
The shortest lifetimes are for the lowest $n>n_\rmc$ and lowest $l$ since the latter can
radiate $(n\rightarrow n')$ to the lowest possible $n'$-states. Thus, the final result is 
relatively insensitive to the range of $n>n_\rmc$ considered, provided that there no other
magnets resulting-in significant cut-offs.

A more elaborate, experimental set-up dependent, approach is described by Schippers \etal (2001)
which takes account of the cooler geometry and the position of the various magnets and their fields as they
impinge upon the recombined ions along their path to ultimate survival to be counted as recombined, or not.
In addition, field ionization rates, due to Damburg and Kolosov (1979), are calculated explicitly.

\subsection{Application to Maxwellian plasmas}
The usual expression for the Maxwellian partial DR rate coefficient (e.g. Badnell \etal 2003) can 
be obtained simply from the corresponding integrated DR cross section, given by equation (\ref{pdri}):
\begin{eqnarray}
\label{maxe}
\alpha^{j}_{f\nu}(T)&=&\left({4\pi a^2_0 I_\rmH \over kT}\right)^{3/2}
{E_\rmc \over \left(2\pi a_0 I_\rmH \right)^{2}\tau_0}\hat{\sigma}^{j}_{f\nu}(E_\rmc)
\exp\left({-E_\rmc \over kT}\right)\,,
\end{eqnarray}
where $(4\pi a^2_0)^{3/2}=6.6011\times 10^{-24}$ cm$^3$.
Trivially, it can also be determined from the energy-averaged DR cross section, $\bar{\sigma}$, 
for $kT>>\Delta E$, on substituting for $\hat{\sigma}$ in equation (\ref{maxe}) from 
equation (\ref{bine}).

Total DR-plus-RR rate coefficients are required for plasma modelling.
We determine the RR contribution also using {\sc autostructure}, following 
Badnell and Seaton (2003) and Badnell (2006).

\subsubsection{Fits to totals.}
It is convenient often for modelling purposes to fit the total (Maxwellian) DR rate coefficient,
$\alpha^{\rm DR}_\nu(T)$, to the following functional form:
\begin{eqnarray}
\alpha^{\rm DR}_\nu(T)&=&T^{-3/2}\sum_i c_i \exp\left({-E_i \over T}\right)\,,
\label{DRfit}
\end{eqnarray}
where the $E_i$ are in the units of temperature, $T$, (eV or K) and the units of $c_i$ are 
then cm$^3 \rms^{-1}$[eV or K]$^{3/2}$.

The same is also true for RR, for which we use the usual (Verner and Ferland 1996) functional form:
\begin{eqnarray}
\fl
\alpha^{\rm RR}_\nu(T)&=&A\left[ \left({T\over T_0}\right)^{1/2} 
\left(1+\left({T\over T_0}\right)^{1/2}\right)^{1-B} 
\left(1+\left({T\over T_1}\right)^{1/2}\right)^{1+B} 
\right]^{-1}\,,
\label{RRfit}
\end{eqnarray}
where $T_0$, $T_1$ are in the units of temperature (eV or K) and the units of $A$ are 
cm$^3 \rms^{-1}$,  while $B$ is dimensionless. A more accurate representation (Gu 2003), 
especially for low-charge ions, replaces $B$ as

\begin{eqnarray}
B\rightarrow B+C\exp\left(- {T_2\over T}\right)\,,
\end{eqnarray}
where, again, $C$ is dimensionless and $T_2$ has the units of temperature.

\section{The Fe$^{13+}$ target}
\label{calc}
The DR reactions which we take account of are defined by the $N$-electron target
configuration interaction expansion which we use. All possible $(N+1)$-electron configurations
are then constructed by adding a continuum or bound orbital to them. All possible
autoionization and (electric dipole) radiative rates are determined from these
configurations, and are applied subsequently so as to determine partial and total
DR cross sections, following the theory of Section \ref{thy}.

It is still the case, for Fe$^{13+}$ at least, that it is convenient and meaningful
to consider separately the $\Delta n=0$ and $\Delta n=1$ core-excitation contributions to DR
since, as we shall see, our highest $n=3$ target level lies below our lowest $n=4$ level.
Indeed, `$\Delta n=0$' DR completely dominates over $\Delta n=1$ at photoionized plasma temperatures,
and can be expected to be the largest contribution too at collisional plasma temperatures, at
least where Fe$^{13+}$ is normally abundant (Mazzotta \etal 1998, Bryans \etal 2006). This separation enables us
to restrict the sum over Rydberg states to $n=1000$, $l=15$ and to $n=100$, $l=5$, for $\Delta n=0$ 
and $\Delta n=1$, respectively.

\subsection{$\Delta n=0$}
We consider two different target configuration interaction expansions because carrying-out 
two DR calculations enables us to contrast the
level of accuracy/difference in the $N$-electron targets with that of differences between
theory and experiment for cross sections and assess the accuracy of total rate coefficients, 
i.e. quantify the uncertainty in the $(N+1)$-electron problem.

We define a Basis A consisting of the following configurations (assuming a closed Ne-like core):
\begin{eqnarray}
1: 3\rms^2 3\rmp\,, \quad\quad 2: 3\rms 3\rmp^2\,, \quad 3: 3\rms^2 3\rmd\,, \quad 4: 3\rmp^3\,,\nonumber \\ 
5: 3\rms 3\rmp 3\rmd\,, \quad\,\, 6: 3\rmp^2 3\rmd\,, \quad 7: 3\rms 3\rmd^2\, \nonumber
\end{eqnarray}
and a Basis B, which consists of Basis A plus
\begin{eqnarray}
8: 3\rmp 3\rmd^2\,. \nonumber
\end{eqnarray}
Thus, $3\rmd^3$ is the only configuration from the $n=3$ complex which is omitted, by Basis B.
Configurations 1 -- 3, plus 5, form the minimal set which allows for all ($\Delta n=0$) one-electron 
promotions during the dielectronic capture process from the ground configuration. Configuration 4 mixes
strongly with 5, whilst 6 and 7 (which are strongly mixed themselves) provide the leading even parity
configuration interaction. Configuration 8 (Basis B) provides a check on that for the odd parity.

Basis A gives rise to 37 target terms whilst Basis B gives rise to 56 terms.
In both cases the radial functions were determined using the Slater-Type-orbital model
potential of Burgess \etal (1989). The (3s, 3p, 3d) radial scaling parameters, $\lambda_{nl}$, were determined by 
minimizing the equally weighted sum of eigenenergies of the 18 lowest terms, which correspond to all of those
which arise from the first 5 configurations of the basis expansions. For Basis A: $\lambda_{3\rms}=0.93173$, 
$\lambda_{3\rmp}=0.99255$ and $\lambda_{3\rmd}=0.89006$. For Basis B: $\lambda_{3\rms}=0.94849$, 
$\lambda_{3\rmp}=1.02316$ and $\lambda_{3\rmd}=0.86884$.
All other radial scaling parameters were taken to be unity.

Basis A gives rise to 84 target levels whilst Basis B gives rise to 129 levels.
Our Breit--Pauli calculations include the one-body non-fine-structure and fine-structure operators,
including the effective one-body Blume and Watson operator for the mutual-spin-orbit and spin-other-orbit 
interactions between valence electrons and the Ne-like closed shells (Badnell 1997). The effect of the
two-body fine-structure operators representing interactions between valence electrons (including
spin-spin now) is small --- of the order $10^{-4}$ Ry --- and since they are time consuming to
determine in the DR calculation we omit them, along with the two-body non-fine-structure operators
which are of the same order effect.

\begin{table}
\caption{Level energies (Ry) for Fe$^{13+}$.\label{tab1}}
\footnotesize
\begin{tabular}{rrrrrrrrr}
\br
Level & Config. & $(2S+1)^\rma$ & $L$ & $2J$ & Basis A$^\rmb$& Basis B$^\rmb$  & Basis 2$^\rmc$ & Observed$^\rmd$ \\
\mr
1   &     1  &     $-$2  & 1  &  1  &	     0.00000    &    0.00000    &  0.00000    &    0.00000	   \\
2   &     1  &     $-$2  & 1  &  3  &	     0.16012    &    0.15817    &  0.16850    &    0.17180	   \\
3   &     2  &      4  & 1  &  1  &	     2.01096    &    2.02837    &  2.02729    &    2.05139	   \\
4   &     2  &      4  & 1  &  3  &	     2.07565    &    2.09213    &  2.09568    &    2.12133	   \\
5   &     2  &      4  & 1  &  5  &	     2.15915    &    2.17464    &  2.18229    &    2.20879	   \\
6   &     2  &      2  & 2  &  3  &	     2.71144    &    2.72742    &  2.73082    &    2.72689	   \\
7   &     2  &      2  & 2  &  5  &	     2.72878    &    2.74444    &  2.74989    &    2.74719	   \\
8   &     2  &      2  & 0  &  1  &	     3.32805    &    3.33949    &  3.36257    &    3.32333	   \\
9   &     2  &      2  & 1  &  1  &	     3.54992    &    3.56180    &  3.58371    &    3.54036	   \\
10  &     2  &      2  & 1  &  3  &	     3.62333    &    3.63506    &  3.65713    &    3.61328	   \\
11  &     3  &      2  & 2  &  3  &	     4.36990    &    4.37479    &  4.38676    &    4.31233	   \\
12  &     3  &      2  & 2  &  5  &	     4.39202    &    4.39760    &  4.40889    &    4.33036	   \\
13  &     4  &     $-$2  & 2  &  3  &	     5.25986    &    5.25043    &  5.25464    &    5.25239	   \\
14  &     4  &     $-$2  & 2  &  5  &	     5.28824    &    5.28126    &  5.28780    &    5.28747	   \\
15  &     4  &     $-$4  & 0  &  3  &	     5.41728    &    5.35636    &  5.37640    &    5.36738	   \\
16  &     5  &     $-$4  & 3  &  3  &	     5.86740    &    5.87474    &  5.87685    &		           \\
17  &     4  &     $-$2  & 1  &  1  &	     5.94570    &    5.86487    &  5.88362    &    5.85316	   \\
18  &     4  &     $-$2  & 1  &  3  &	     5.96395    &    5.89096    &  5.91009    &    5.88140	   \\
19  &     5  &     $-$4  & 3  &  5  &	     5.90286    &    5.91136    &  5.91332    &    5.88668	   \\
20  &     5  &     $-$4  & 3  &  7  &	     5.95450    &    5.96289    &  5.96590    &    5.94097	   \\
21  &     5  &     $-$4  & 3  &  9  &	     6.02572    &    6.03378    &  6.03878    &    6.01676	   \\
22  &     5  &     $-$4  & 1  &  5  &	     6.33030    &    6.31638    &  6.33538    &    6.29051	   \\
23  &     5  &     $-$4  & 2  &  3  &	     6.34870    &    6.33224    &  6.35618    &    6.31200	   \\
24  &     5  &     $-$4  & 2  &  1  &	     6.36253    &    6.34247    &  6.37066    &    6.32572	   \\
25  &     5  &     $-$4  & 2  &  7  &	     6.44827    &    6.42579    &  6.45619    &    6.40979	   \\
26  &     5  &     $-$4  & 1  &  1  &	     6.43441    &    6.42258    &  6.44598    &    6.41304	   \\
27  &     5  &     $-$4  & 2  &  5  &	     6.44913    &    6.42896    &  6.45759    &    6.41636	   \\
28  &     5  &     $-$4  & 1  &  3  &	     6.44322    &    6.42749    &  6.45342    &    6.41722	   \\
29  &     5  &     $-$2  & 2  &  3  &	     6.67429    &    6.56724    &  6.59249    &    6.53556	   \\
30  &     5  &     $-$2  & 2  &  5  &	     6.67611    &    6.57330    &  6.59785    &    6.54163	   \\
31  &     5  &     $-$2  & 3  &  5  &	     6.90910    &    6.86220    &  6.88292    &    6.78862	   \\
32  &     5  &     $-$2  & 3  &  7  &	     7.03625    &    6.98836    &  7.01333    &    6.92393	   \\
33  &     5  &     $-$2  & 1  &  3  &	     7.49502    &    7.41984    &  7.46965    &    7.35495	   \\
34  &     5  &     $-$2  & 1  &  1  &	     7.54717    &    7.48992    &  7.54320    &		           \\
35  &     5  &     $-$2  & 3  &  7  &	     7.67057    &    7.55001    &  7.57352    &    7.45046	   \\
36  &     5  &     $-$2  & 3  &  5  &	     7.69403    &    7.57286    &  7.59829    &    7.47787	   \\
37  &     5  &     $-$2  & 1  &  1  &	     7.97268    &    7.75185    &  7.79103    &    7.65001	   \\
38  &     5  &     $-$2  & 2  &  3  &	     7.92298    &    7.74867    &  7.80163    &    7.66171	   \\
39  &     5  &     $-$2  & 1  &  3  &	     8.00254    &    7.78798    &  7.82900    &    7.68796         \\
40  &     5  &     $-$2  & 2  &  5  &	     7.95856    &    7.77661    &  7.83024    &    7.69544	   \\
 				   
\br
\end{tabular}
\newline
$^{\rma} >0 $ denotes even parity, $<0$ odd parity.
\newline
$^{\rmb}$This work.
\newline
$^{\rmc}$Storey \etal (2000).
\newline
$^{\rmd}$NIST (2006).
\end{table}

In Table \ref{tab1} we compare our lowest 40 calculated level energies obtained from using Bases A and B with those 
obtained from the NIST (2006) database and those calculated with {\sc superstructure} by Storey \etal (2000)
using the Thomas--Fermi model potential. These levels are all of those which arise from the lowest 5 configurations, 
i.e, it includes all levels which contribute to $\Delta n=0$ DR in the absence of configuration mixing. 
We note a distinct improvement in the agreement with the results of (Basis 2 of) Storey \etal (2000), and 
with the observed energies, on going from Basis A to Basis B.
Basis 2 of Storey \etal (2000) included all configurations from the $n=3$ complex as well as $n=4$ configurations 
of the form $3\rms^2 4l$ and $3\rms 3\rmp 4l$, for $l=0-3$.
We note little improvement in the agreement with the observed energies resulting from the use
by Storey \etal of their larger target basis 2, compared to Basis B.
Some high-lying levels (37--40) are now in observed order but, on the other hand, many of the levels of
configuration 5 ($3\rms3\rmp3\rmd$) are distinctly higher, compared to the observed, than are those from Basis B.

Nevertheless, differences of up to 0.07 Ry (mostly up to 0.03 Ry for Basis B) between the calculated and 
observed low-lying level energies
means that it is important to use the observed target energies to position the DR resonances,
so as to eliminate sensitivity to the exponential factor in equation (\ref{maxe}) at photoionized plasma 
temperatures. This is done simply by moving each $(N+1)$-electron autoionizing level by the difference
between the calculated and observed excitation energies between the initial and parent
$N$-electron level energies. 

\begin{table}
\caption{Radiative transition rates, $A^{{\rm r}}(\rms^{-1})$, for Fe$^{13+}$.\label{tab2}}
\footnotesize
\begin{tabular}{rrrrrrrrrrr}
\br
$j$& $f$ &  Basis A$^\rma$& Basis B$^\rma$  &Basis 2$^{\star\rmb}$& &$j$& $f$  & Basis A$^\rma$  & Basis B$^\rma$& Basis 2$^{\star\rmb}$ \\
\mr
3   &   1$-$2  &  3.25(07)$^\rmc$  &  3.25(07)	&  3.66(07)  &    &   25  &	7  &  2.75(08)    &  2.53(08)	 &  3.26(08)	    \\
4   &   1$-$2  &  6.07(06)  &  6.28(06)	&  6.78(06)  &    &   26  &	4  &  3.09(10)    &  2.97(10)	 &  2.79(10)	    \\
5   &     2  &  2.19(07)  &  2.28(07)	&  2.65(07)  &    &   26  &	6  &  3.94(07)    &  3.67(07)	 &  4.29(07)	    \\
6   &   1$-$2  &  2.56(09)  &  2.73(09)	&  2.46(09)  &    &   27  &   4$-$5  &  4.08(10)    &  3.92(10)	 &  3.71(10)	    \\
7   &     2  &  2.07(09)  &  2.23(09)	&  1.91(09)  &    &   27  &   6$-$7  &  4.61(07)    &  1.27(08)	 &  1.82(08)	    \\
8   &   1$-$2  &  1.95(10)  &  1.95(10)	&  1.89(10)  &    &   28  &   4$-$5  &  3.59(10)    &  3.40(10)	 &  3.19(10)	    \\
9   &   1$-$2  &  3.80(10)  &  3.90(10)	&  3.43(10)  &    &   28  &   6$-$7  &  1.14(08)    &  1.41(08)	 &  1.93(08)	    \\
10  &   1$-$2  &  4.36(10)  &  4.47(10)	&  4.05(10)  &    &   29  &   3$-$4  &  3.29(08)    &  4.33(08)	 &  5.38(08)	    \\
11  &   1$-$2  &  4.71(10)  &  4.65(10)	&  4.36(10)  &    &   29  &   6$-$7  &  3.95(10)    &  3.54(10)	 &  3.33(10)	    \\
12  &     2  &  4.32(10)  &  4.28(10)	&  3.97(10)  &    &   29  &	8  &  3.93(08)    &  2.87(08)	 &  3.49(08)	    \\
13  &   3$-$5  &  4.85(08)  &  1.14(09)	&  1.24(09)  &    &   29  &  9$-$10  &  1.76(09)    &  1.49(09)	 &  1.36(09)	    \\
13  &   6$-$7  &  3.27(09)  &  3.50(09)	&  2.99(09)  &    &   29  &    11  &  5.28(08)    &  4.82(08)	 &  4.07(08)	    \\
13  &     8  &  3.08(08)  &  3.41(08)	&  3.80(08)  &    &   30  &   4$-$5  &  3.57(08)    &  5.30(08)	 &  6.09(08)	    \\
13  &  9$-$10  &  5.97(08)  &  6.04(08)	&  5.00(08)  &    &   30  &   6$-$7  &  3.96(10)    &  3.51(10)	 &  3.28(10)	    \\
13  &    11  &  7.11(06)  &  6.13(06)	&  6.51(06)  &    &   30  &    10  &  1.20(09)    &  9.72(08)	 &  8.45(08)	    \\
14  &     5  &  7.58(07)  &  8.23(07)	&  9.50(07)  &    &   30  & 11$-$12  &  5.96(08)    &  5.21(08)	 &  4.45(08)	    \\
14  &   6$-$7  &  3.46(09)  &  3.76(09)	&  3.25(09)  &    &   31  &   4$-$5  &  9.95(07)    &  8.55(08)	 &  1.10(08)	    \\
14  &    10  &  7.02(08)  &  7.30(08)	&  6.67(08)  &    &   31  &   6$-$7  &  1.96(10)    &  1.71(10)	 &  1.55(10)	    \\
14  &    12  &  7.47(06)  &  6.82(06)	&  7.71(06)  &    &   31  & 11$-$12  &  1.45(09)    &  1.80(09)	 &  1.54(09)	    \\
15  &   3$-$5  &  4.06(10)  &  3.78(10)	&  3.41(10)  &    &   32  &	5  &  4.04(08)    &  3.68(08)	 &  4.36(08)	    \\
15  &     6  &  4.54(07)  &  1.10(08)	&  1.09(08)  &    &   32  &	7  &  2.10(10)    &  1.81(10)	 &  1.62(10)	    \\
15  &    10  &  5.18(07)  &  6.00(07)	&  6.45(07)  &    &   32  &    12  &  1.55(09)    &  2.03(09)	 &  1.95(09)	    \\
16  &   3$-$4  &  9.87(07)  &  1.85(08)	&  1.53(08)  &    &   33  &	3  &  2.82(08)    &  2.56(08)	 &  2.87(08)	    \\
16  &   6$-$7  &  1.52(08)  &  1.16(09)	&  4.53(08)  &    &   33  &	8  &  4.69(10)    &  4.13(10)	 &  4.03(10)	    \\
16  &     8  &  6.11(06)  &  1.92(08)	&  5.54(07)  &    &   33  &  9$-$10  &  1.00(10)    &  1.19(10)	 &  1.00(10)	    \\
16  &  9$-$10  &  4.11(07)  &  2.34(08)	&  5.75(07)  &    &   33  & 11$-$12  &  6.70(08)    &  8.76(08)	 &  9.77(08)	    \\
16  &    11  &  1.51(06)  &  4.13(06)	&  3.24(06)  &    &   34  &	3  &  1.28(08)    &  1.18(08)	 &  1.28(08)	    \\
17  &   3$-$4  &  3.81(07)  &  4.03(07)   &  4.17(07)  &    &   34  &	6  &  3.81(08)    &  9.49(07)	 &  2.54(08)	    \\
17  &     6  &  1.44(10)  &  1.35(10)   &  1.21(10)  &    &   34  &	8  &  2.11(10)    &  1.60(10)	 &  1.44(10)	    \\
17  &     8  &  1.56(08)  &  2.16(08)   &  9.18(07)  &    &   34  &  9$-$10  &  3.02(10)    &  3.30(10)	 &  3.09(10)	    \\
17  &  9$-$10  &  4.19(09)  &  4.47(09)   &  3.90(09)  &    &   34  &    11  &  1.13(09)    &  1.01(08)	 &  2.80(08)	    \\
18  &   3$-$5  &  7.87(08)  &  6.98(08)   &  8.37(08)  &    &   35  &	5  &  2.80(08)    &  2.60(08)	 &  3.23(08)	    \\
18  &   6$-$7  &  1.29(10)  &  1.10(10)   &  1.04(10)  &    &   35  &	7  &  2.84(10)    &  2.90(10)	 &  2.81(10)	    \\
18  &     8  &  1.84(09)  &  1.90(09)   &  1.69(09)  &    &   35  &    12  &  2.98(10)    &  2.61(10)	 &  2.22(10)	    \\
18  &    10  &  2.85(09)  &  2.80(09)   &  2.56(09)  &    &   36  &   6$-$7  &  2.97(10)    &  2.98(10)	 &  2.87(10)	    \\
19  &   4$-$5  &  2.06(08)  &  2.21(08)   &  2.48(08)  &    &   36  &    10  &  2.61(08)    &  4.70(08)	 &  3.76(08)	    \\
19  &   6$-$7  &  5.98(07)  &  7.70(07)   &  8.57(07)  &    &   36  & 11$-$12  &  3.10(10)    &  2.73(10)	 &  2.24(10)	    \\
19  &    10  &  6.38(05)  &  6.57(05)   &  5.65(05)  &    &   37  &	8  &  1.43(10)    &  1.78(10)	 &  1.57(10)	    \\
19  &    11  &  3.42(06)  &  4.22(06)   &  4.99(06)  &    &   37  &  9$-$10  &  3.22(10)    &  1.98(10)	 &  1.98(10)	    \\
20  &     5  &  2.38(08)  &  2.57(08)   &  2.94(08)  &    &   37  &    11  &  3.66(10)    &  3.26(10)	 &  2.75(10)	    \\
20  &     7  &  5.21(05)  &  5.28(05)   &  1.06(06)  &    &   38  &   6$-$7  &  2.67(08)    &  2.38(08)	 &  2.78(08)	    \\
20  &    12  &  4.24(06)  &  5.05(06)   &  6.25(06)  &    &   38  &	8  &  3.20(09)    &  1.11(09)	 &  8.17(07)	    \\
22  &   4$-$5  &  3.05(10)  &  3.00(10)   &  2.82(10)  &    &   38  &  9$-$10  &  6.41(10)    &  5.92(10)	 &  4.53(10)	    \\
22  &	6$-$7  &  4.68(08)  &  9.44(08)   &  1.13(09)  &    &   38  & 11$-$12  &  2.56(10)    &  2.15(10)	 &  2.21(10)	    \\
22  &	 10  &  2.35(07)  &  3.30(07)   &  3.73(07)  &    &   39  &   6$-$7  &  4.02(08)    &  4.76(08)	 &  4.65(08)	    \\
22  &	 12  &  1.71(07)  &  2.73(07)   &  3.00(07)  &    &   39  &	8  &  2.65(09)    &  6.13(09)	 &  6.13(09)	    \\
23  &	3$-$5  &  3.74(10)  &  3.74(10)   &  3.55(10)  &    &   39  &  9$-$10  &  3.99(10)    &  2.82(10)	 &  3.56(10)	    \\
23  &	  6  &  2.48(08)  &  3.86(08)   &  5.04(08)  &    &   39  & 11$-$12  &  3.85(10)    &  3.39(10)	 &  2.65(10)	    \\
24  &	3$-$4  &  4.40(10)  &  4.35(10)   &  4.05(10)  &    &   40  &    10  &  7.07(10)    &  6.25(10)	 &  5.81(10)	    \\
25  &	  5  &  4.43(10)  &  4.29(10)   &  4.04(10)  &    &   40  & 11$-$12  &  2.42(10)    &  2.13(10)	 &  1.95(10)	    \\
\br
\end{tabular}
\newline
$^{\rma}$This work.
\newline
$^{\rmb}$Extended Basis 2, Storey \etal (2000).
\newline
$^{\rmc}$3.25(07) denotes $3.25\times 10^7$.
\end{table}

Generally, $A^{{\rm r}} << A^{{\rm a}}$ for $\Delta n=0$ DR and so $\hat{\sigma}\propto A^{{\rm r}}$ 
(see equation \ref{pdri}). Excepting DR via the fine-structure core-excitation ($3\rmp_{1/2} - 3\rmp_{3/2}$),
$A^{{\rm r}}$ is dominated by the inner-electron (dipole) radiative rate. Thus, it is instructive to study
radiative rates for Fe$^{13+}$ in some detail.
In Table \ref{tab2} we compare our radiative rates obtained from using
Bases A and B with those determined by Storey \etal (2000) from the `extended' Basis 2 
of Storey \etal (1996). This `extended' Basis 2 includes the configurations of Basis 2 (Storey \etal 2000)
but adds further configurations involving $n=4$ orbitals. In addition, all of the $n=4$ orbitals are now 
pseudo-orbitals (they were physical in the vanilla Basis 2) --- see Storey \etal (1996) for further details.
As far as (total) DR cross sections are concerned, the distribution of radiative rates over the final states
is irrelevant, in general, so long as they are all bound. Thus, we have summed over the fine-structure levels of
the lower term to make the comparison shown in Table \ref{tab2}.

Overall, we observe no drastic difference between the results obtained on using Basis A and B compared to those of
Storey \etal (2000). The results of Bases B tend to be closer to those of Storey \etal (2000), than Basis A,
for the strong radiative rates ($\sim 10^{10}\,\, \rms^{-1}$), especially from configuration 5 ($3\rms 3\rmp 3\rmd$).
A general point is illustrated by the results for the spin-quartet level 16. This level mixes with the nearby doublet 
level 18 (both $J=3/2$, odd parity). Consequently, rates from level 16 to lower-lying spin-doublets  are
very sensitive to the precise mixing. Of course, as far as DR is concerned, if the autoionization rates
associated with these parents are such that $A^{{\rm a}} >> A^{{\rm r}}$ then the DR cross section is
simply redistributed from one peak to another, the parents being less than 0.5eV apart.

\subsection{$\Delta n=1$}
$\Delta n=1$ core-excitation contributions to DR come into play at high temperatures, i.e., in electron collision
dominated plasmas. Like the case of $1-2$ core-excitations in Li-, Be- and B-like ions, we expect the contribution 
from `inner-shell' $2-3$ core-excitations to rapidly decrease as we progress through Na-, Mg- and Al-like due
to the increasing number of core-rearrangement autoionization channels.
We consider both $2-3$ and $3-4$ $\Delta n=1$ core-excitations so as to get a precise assessment of their
relative importance and an indication of whether, or not, $2-3$ core-excitations need to be considered beyond Al-like.
The $\Delta n=1$ contribution to a Maxwell rate coefficient is not sensitive
to resonance positions, and so the {\it ab initio} calculated energies were used throughout.

\subsubsection{$3-4$.}
Again, we used two different configuration basis sets for $3-4$ core-excitations so as to gain insight into the 
uncertainty of the theoretical results. The first (Basis C) consisted of Basis A, plus $3\rms^2 4l\,, 3\rms 3\rmp 4l\,,
3\rmp^2 4l$, for $l=0-3$. These 19 configurations give rise to 250 target levels. The scaling parameters for
the Slater-Type-Orbital model potential were determined from a subset of these configurations: $3\rms^2 nl$, 
for $n=3, 4$ and all $l$. The reason for this is to ensure that the optimization procedure for the $n=4$ orbitals
was tightly linked to the $n=4$ term energies. The $3\rmp$ and $3\rmd$ scaling parameters were determined
first by minimizing on the lowest two terms. These were then fixed and the $4l$ scaling parameters determined
by minimizing the equally weighted sum of $n=4$ term energies (still in the presence of the $n=3$ states).
The result: $\lambda_{3\rmp}=0.69556$, $\lambda_{3\rmd}=0.71607$, $\lambda_{4\rms}=0.79659$, $\lambda_{4\rmp}=0.7903$, 
$\lambda_{4\rmd}=0.7821$, and $\lambda_{4\rmf}=0.8563$. The $3\rms$ parameter is not well determined by such a
procedure, on the other hand, opening-up the $3\rms$ sub-shell would not be consistent with the procedure for
the other orbitals. Hence, we simply set $\lambda_{3\rms}=0.7$ for consistency with the other scaling parameters.
Such a procedure is optimum for $3\rmp \rightarrow 4l$ promotions, which can be expected to dominate the $3-4$
core-excitations.
 
The second basis that we use (Basis D) is the scattering target basis used by Storey \etal (1996) for the electron-impact
excitation of Fe$^{13+}$. It includes all configurations belonging to the $n=3$ complex, plus $3\rms^2 4l\,, 
3\rms 3\rmp 4l\,$, for $l=0-3$. Thus, it includes the $3\rmp3\rmd^2$ and $3\rmd^3$ configurations omitted
by Basis C but excludes the $3\rmp^2 4l$ configurations. The focus of their work was excitation within the
$n=3$ complex, including resonances attached to $n=4$. Resonances attached to $n=4$ are our primary concern here.
Storey \etal (1996) used $nl$-dependent Thomas--Fermi model potentials, and the relevant values of the scaling
parameters are listed in their Table 1. There are 227 levels associated with this 17 configuration basis.

\begin{table}
\caption{Some $n=4$ level energies (Ry) for Fe$^{13+}$, relative to the ground level.\label{tab3}}
\begin{indented}
\item[]\begin{tabular}{rrrrrrr}
\br
 Config. & $(2S+1)^\rma$ & $L$ & $2J$ & Basis C$^\rmb$& Basis D$^\rmc$   & Observed$^\rmd$ \\
\mr
  4\rms  &     2  & 0  &  1  &	     12.9023    &    13.2113      &    13.0769         \\
  4\rmp  &    $-$2  & 1  &  1  &	     13.9497    &    14.2595      &    14.2963         \\
  4\rmp  &    $-$2  & 1  &  3  &	     14.0114    &    14.3204      &    14.3434         \\
  4\rmd  &     2  & 2  &  3  &	     15.3886    &    15.6841      &    15.4549         \\
  4\rmd  &     2  & 2  &  5  &	     15.4007    &    15.6965      &    15.4668         \\
  4\rmf  &    $-$2  & 3  &  5  &	     16.2484    &    16.5705      &    16.2969         \\
  4\rmf  &    $-$2  & 3  &  7  &	     16.2518    &    16.5738      &    16.2993         \\
\br
\end{tabular}
 \item[]$^{\rma} >0 \equiv$ denotes even parity, $<0$ odd parity.
 \item[]$^{\rmb}$This work.
 \item[]$^{\rmc}$Recalculated from Storey \etal (1996).
 \item[]$^{\rmd}$NIST (2006).
\end{indented}
\end{table}

\begin{table}
\caption{Symmetric oscillator strengths ($gf$) for Fe$^{13+}$.\label{tab4}}
\begin{indented}
\item[]\begin{tabular}{crrrr}
\br
  &    \centre{2}{Basis C$^\rmb$}& \centre{2}{Basis D$^\rmc$} \\
  &               \crule{2}&\crule{2} \\
 Transition  & Length & Velocity & Length & Velocity  \\
\mr
  $3\rmp - 4\rms$  &   0.2043	&    0.2963	  &    0.3707       &	0.3757   \\
  $3\rmp - 4\rmd$  &   0.9886	&    1.3858	  &    1.5037       &	1.5432   \\
  $3\rmd - 4\rmp$  &   0.1766	&    0.2492	  &    0.2670       &	0.2493   \\
  $3\rmd - 4\rmf$  &   6.0729	&    7.3666	  &    7.4848       &	7.2178   \\
\br
\end{tabular}
 \item[]$^{\rmb}$This work.
 \item[]$^{\rmc}$Recalculated from Storey \etal (1996).
\end{indented}
\end{table}

In table \ref{tab3} we compare energies for the $n=4$ levels which result from the dominant $3\rmp \rightarrow 4l$ 
promotions. We note that the $3\rmp \rightarrow 4\rmd,\, 4\rmf$ excitation energies are better represented by Basis C,
while Basis D is somewhat better for the lower $l$. In table \ref{tab4} we compare symmetric oscillator strengths
for the $3l-4l'$ array, we use the LS-coupling values for simplicity. We note 20--30\% differences between the results
for the two bases --- nearly a factor of 2 for $3\rmp - 4\rms$. There is also much closer agreement between the
length and velocity results for Basis D compared to Basis C. Thus, it is of interest to see how this translates
into differences in the DR cross sections. These differences affect not only the radiative stabilization rates
but also the (dipole) dielectronic capture and autoionization rates, including autoionization into excited states.

\subsubsection{$2-3$.}
We considered $2\rmp\rightarrow 3l$ promotions only. The contribution from $2\rms \rightarrow 3l$ is $<5\%$ of the $2\rmp$
by Ne-like ions. We used a target basis which comprised the first 5 configurations of Basis A, plus $2\rmp^5 3\rms^2 3\rmp^2$, 
$2\rmp^5 3\rms^2 3\rmp 3\rmd$, and $2\rmp^5 3\rms 3\rmp^3$; the latter so as to allow for the strong configuration
mixing with the prior. These configurations give rise to 178 target levels. In addition, we now need to allow for 
core-rearrangement  autoionization transitions of the 
form: $2\rmp^5 3l^4 nl' \rightarrow 2\rmp^6 3l^2 nl' + \rme^-$, i.e., where the Rydberg electron is a spectator. 
These transitions strongly suppress DR since they are `additional' autoionization pathways (ones which have no
reverse dielectronic capture process to balance them) and are independent of $n$, while the populating dielectronic 
capture rate scales as $n^{-3}$.

\section{Velocity-convoluted DR cross section results}
\label{res}

In order to make a comparison between our theoretical DR cross sections and those measured
by Schmidt \etal (2006) at TSR, we convolute them according to equation (\ref{exp1}) using the
experimentally determined $kT_\perp=1.2\times 10^{-2}$~eV and $kT_\parallel=9.0\times 10^{-5}$~eV, 
and apply initially a hard cut-off (see equation \ref{ncut}) at $n_\rmc=45$.\footnote{The value of
$n_\rmc=55$ given by Schmidt \etal (2006) is incorrect (Schmidt, private communication).}
It is assumed that there is no significant metastable fraction in the ion beam when it comes to the 
comparison with experiment, and so we consider only DR from the ground level of Fe$^{13+}$.

\subsection{Parental contributions}
Since many Rydberg series
contribute to the final observed `spectrum', we first show results for various core-excitations,
labelled according to parent level (see table \ref{tab1}) or configuration. We also compare results 
obtained using target Bases A and B. Note the caveat, parentage is not a good quantum number.

\begin{figure}
\begin{flushright} 
\psfig{file=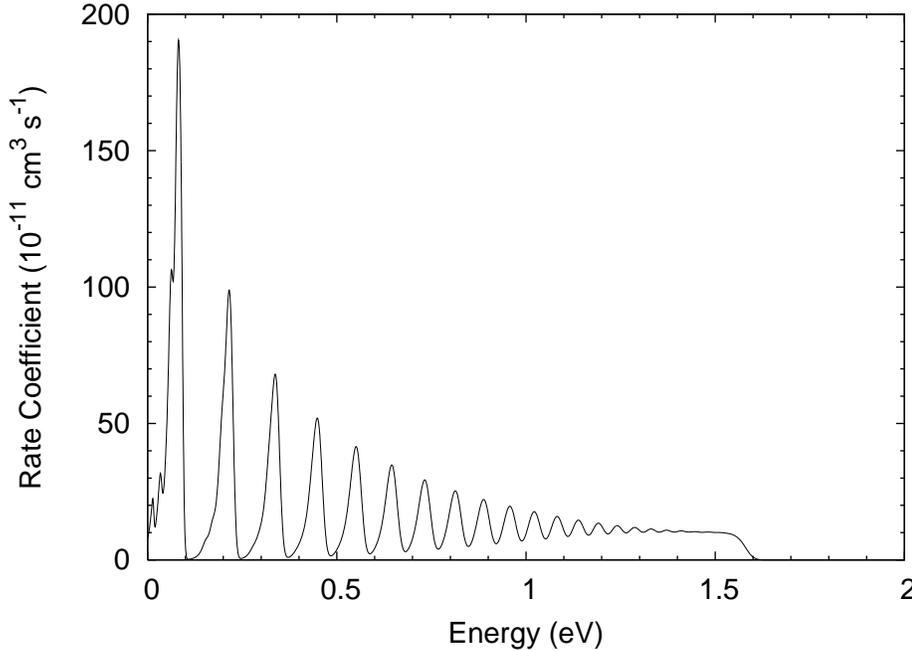}
\end{flushright}
\vspace{-.25in}
\caption{Velocity-convoluted DR cross sections for the fine-structure core-excitation
in the ground term of Fe$^{13+}$.
\label{fig-cf-1}} 
\end{figure}

In Figure \ref{fig-cf-1} we show the convoluted theoretical DR results, which have the dimension
of a rate coefficient, for the fine-structure core-excitation, i.e., parent level 1 to parent level 2,
as listed in Table \ref{tab1}. This opens-up at $n=32$ and falls-off rapidly in $n$, as radiative
stabilization takes place via an outer-electron transition $n\rightarrow n'$, for $n'<32$, and which
is dominated by $n'<10$. The results from Basis A and Basis B are indistinguishable in the figure.

\begin{figure}
\begin{flushright} 
\psfig{file=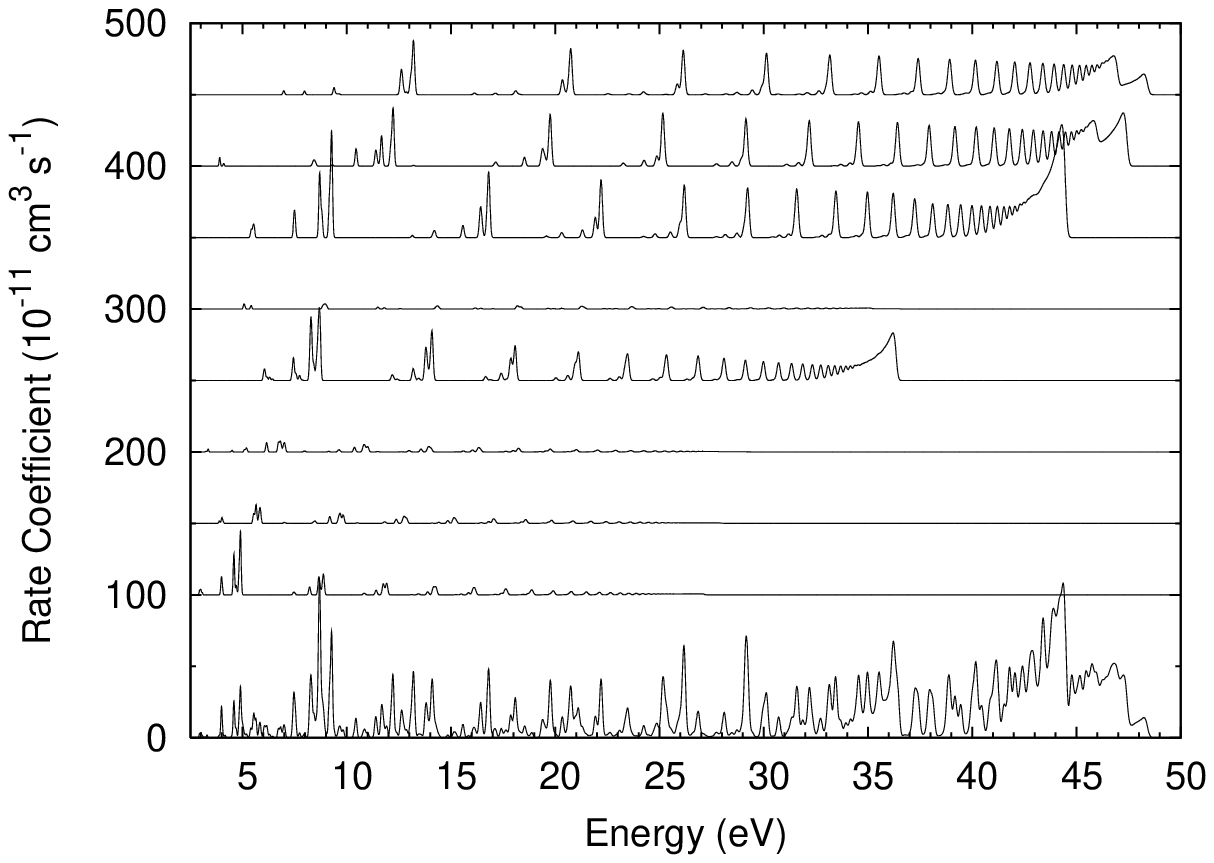}
\end{flushright}
\vspace{-.25in}
\caption{Velocity-convoluted DR cross sections for the $3\rms\rightarrow 3\rmp$
core-excitation to parent configuration 2. Bottom curve, summed-over all parent levels, $i=3-10$.
Offset by $1\times10^{-9} +(i-3)5\times10^{-10}$, the contributions from the individual parent levels, $i$.
\label{fig-cf-2}} 
\end{figure}

In Figure \ref{fig-cf-2} we illustrate the complexity which arises from the $3\rms\rightarrow 3\rmp$
core-excitation to parent configuration 2. The lowest spectrum is the sum of contributions from the 
$^4\rmP_J$, $^2\rmD_J$, $^2\rmS_J$ and $^2\rmP_J$ levels. Offset above are the individual parent level,
$i$, contributions for $i=3-10$. The results from Basis A and Basis B are barely distinguishable in the 
total spectrum.

\begin{figure}
\begin{flushright} 
\psfig{file=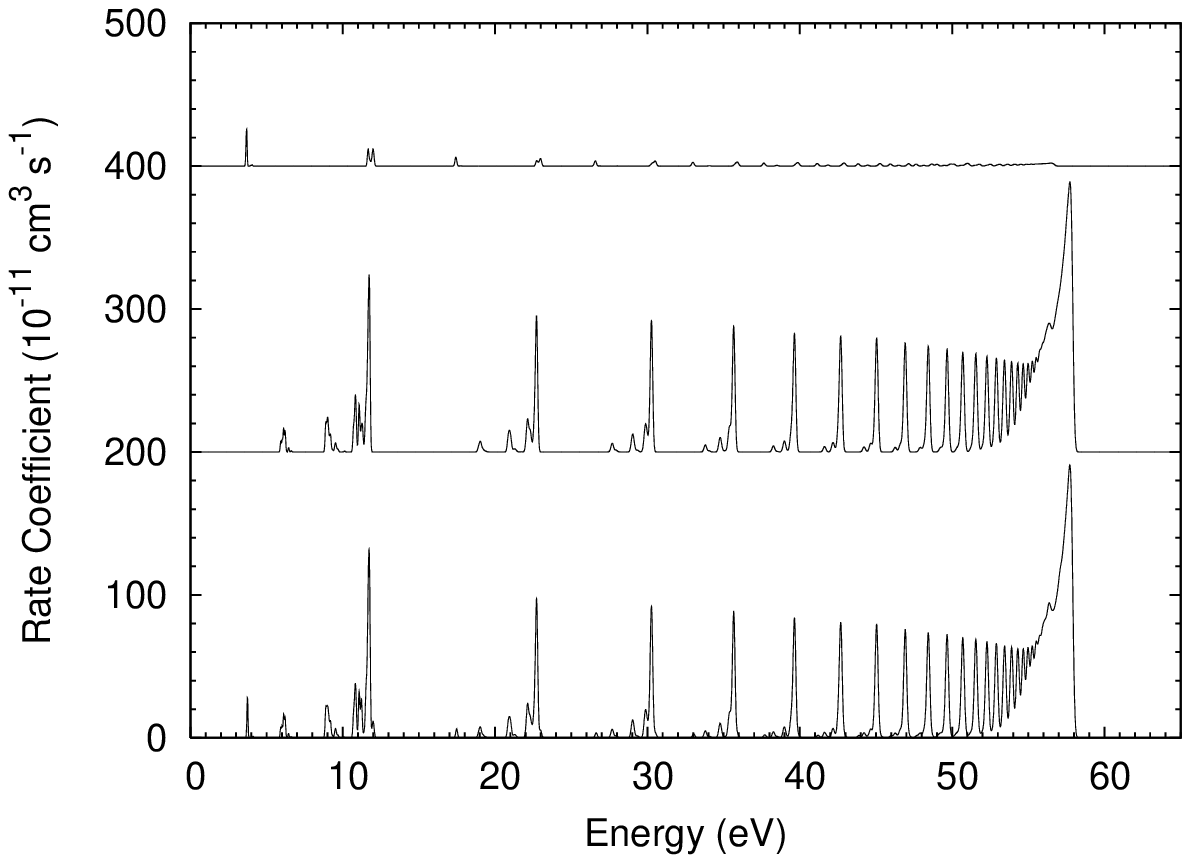}
\end{flushright}
\vspace{-.25in}
\caption{Velocity-convoluted DR cross sections for the $3\rmp\rightarrow 3\rmd$
core-excitation to parent configuration 3. Bottom curve, summed-over all parent levels, $i=11-12$.
Offset by $(i-10)2\times10^{-9}$, the contributions from the individual parent levels, $i$.
\label{fig-cf-3}} 
\end{figure}

In Figure \ref{fig-cf-3} we present results for the $3\rmp \rightarrow 3\rmd$ core-excitation.
There are only two parent levels and the $^2\rmD_{5/2}$ parent level 12 contributes only
weakly as the $J=5/2 \rightarrow 1/2$ core radiative transition is electric dipole forbidden.
Again, the results from Basis A and Basis B are indistinguishable in the figure.

\begin{figure}
\begin{flushright} 
\psfig{file=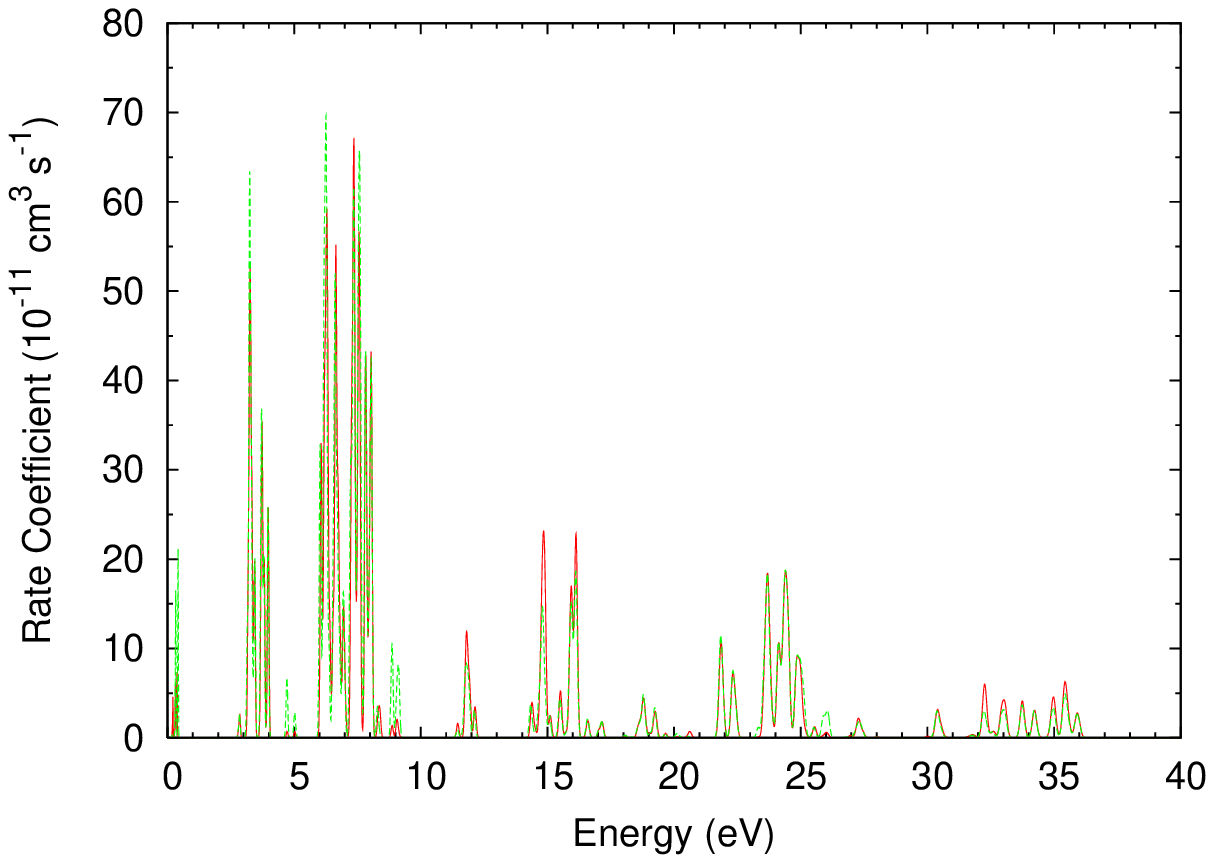}
\end{flushright}
\vspace{-.25in}
\caption{Velocity-convoluted DR cross sections for excitation
of the parent $3\rmp^3$ configuration. Solid (red) curve, Basis A; dashed (green) curve, Basis B.
\label{fig-cf-4}} 
\end{figure}

In figure \ref{fig-cf-4} we compare the results from Basis A and Basis B for excitation
of the parent $3\rmp^3$ configuration 4, summed-over all 5 parent levels. This excitation
only takes place through configuration mixing and we note (i) that it is weaker, especially above
10~eV, and (ii) that we now see some small differences between the results of Basis A and Basis B.
The lowest autoionizing states have $n=6$. Core radiative stabilization is allowed to parent 
configuration 2 via  $3\rmp \rightarrow 3\rms$.
These `recombined' levels then first autoionize at $n=7$  or 8, for the spin-doublet parents.

\begin{figure}
\begin{flushright} 
\psfig{file=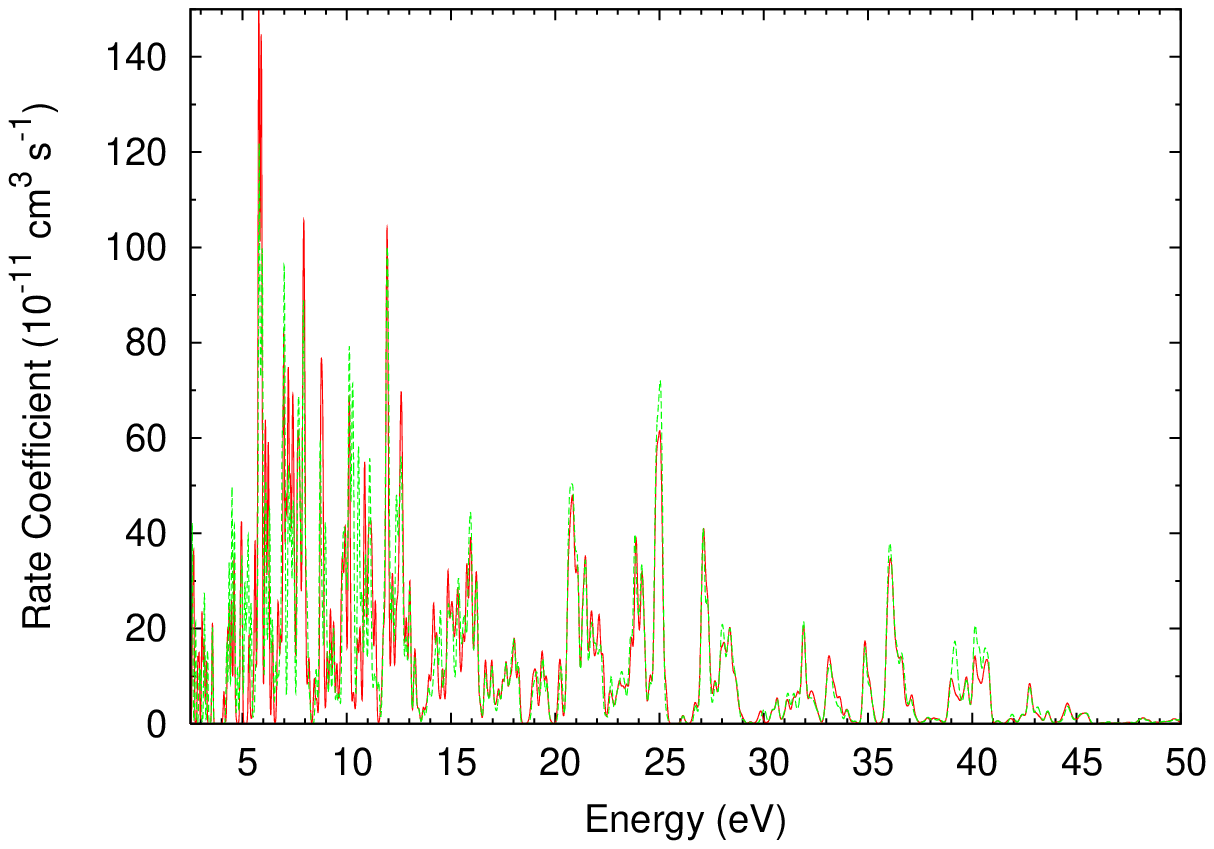}
\end{flushright}
\vspace{-.25in}
\caption{Velocity-convoluted DR cross sections for the $3\rms \rightarrow 3\rmd$
core-excitation of parent configuration 5. Solid (red) curve, Basis A; dashed (green) curve, Basis B.
\label{fig-cf-5}} 
\end{figure}
In figure \ref{fig-cf-5} we compare the results from Basis A and Basis B for the $3\rms \rightarrow 3\rmd$  
core-excitation of parent configuration 5, summed-over all 23 parent levels. Again, there are
small differences, below 15~eV. The lowest autoionizing states have $n=5$ or $6$, depending on the
parent. There are two main core radiative stabilization pathways: $3\rmd \rightarrow 3\rmp$, to
parent configuration 2, and $3\rmp \rightarrow 3\rms$, parent configuration 3. These then first
autoionize at $n=7$ for the latter, and, again, between $n=7$  and 8 for spin-doublet parents of the former.
So, just a few $n$-values contribute strongly, but they are spread out in energy because of the
25~eV spread of levels of configuration 5.

\begin{figure}
\begin{flushright} 
\psfig{file=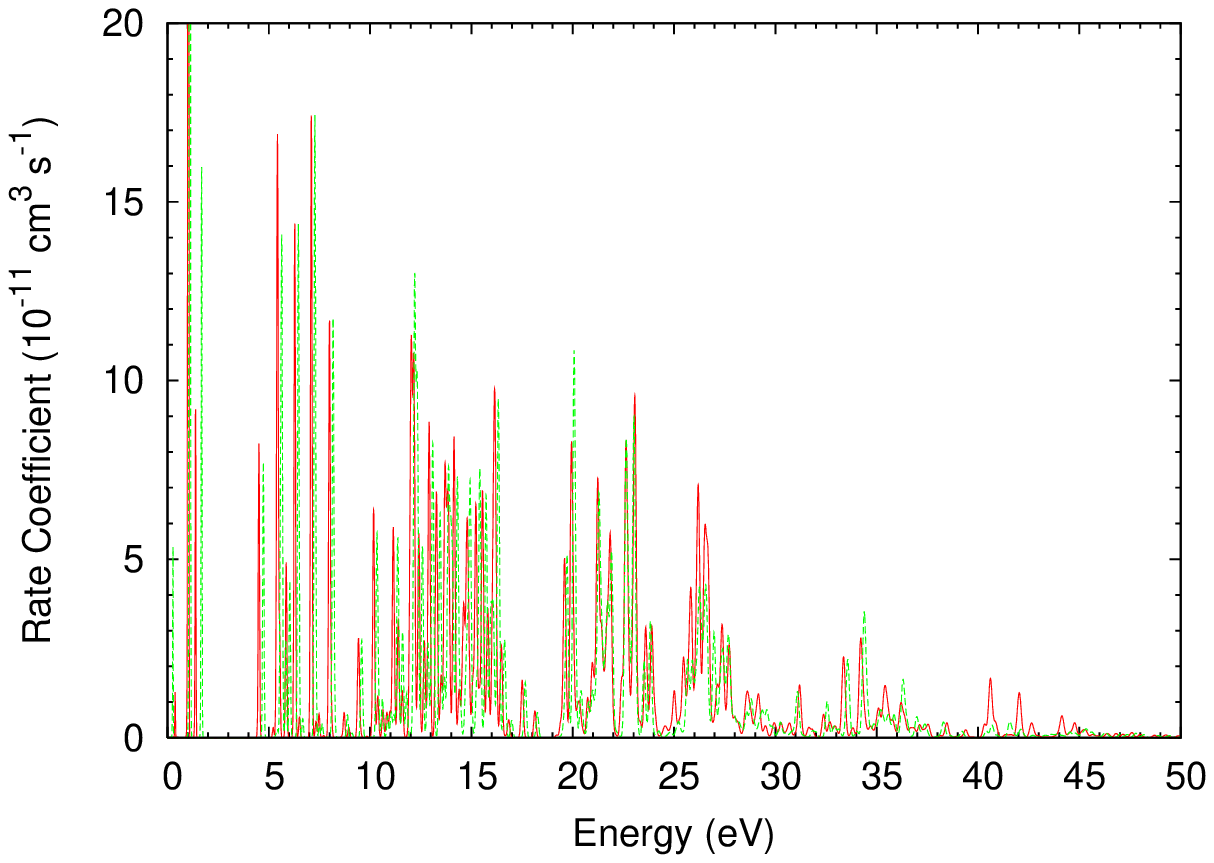}
\end{flushright}
\vspace{-.25in}
\caption{Velocity-convoluted DR cross sections for excitation
of the parent configuration 6. Solid (red) curve, Basis A; dashed (green) curve, Basis B.
\label{fig-cf-6}} 
\end{figure}

\begin{figure}
\begin{flushright} 
\psfig{file=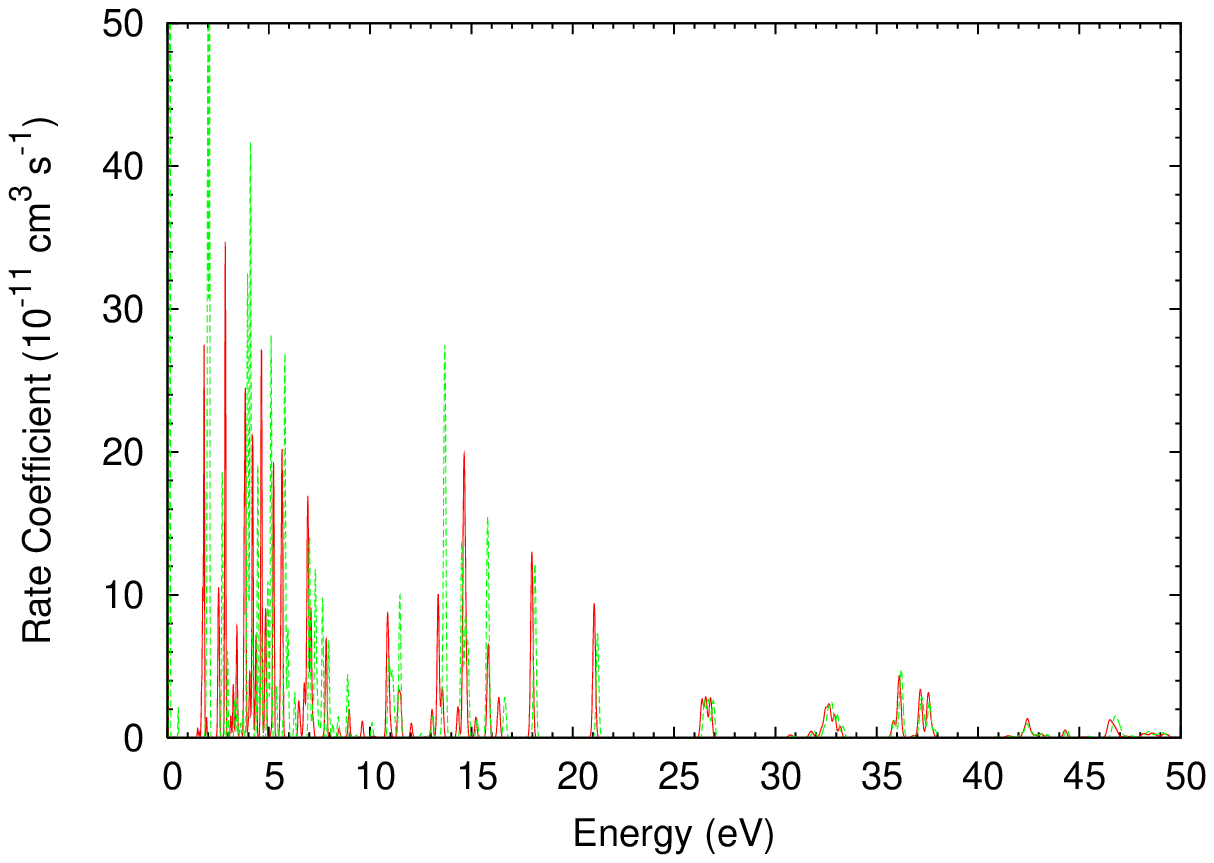}
\end{flushright}
\vspace{-.25in}
\caption{Velocity-convoluted DR cross sections for excitation
of the parent configuration 7. Solid (red) curve, Basis A; dashed (green) curve, Basis B.
\label{fig-cf-7}} 
\end{figure}
\begin{figure}
\begin{flushright} 
\psfig{file=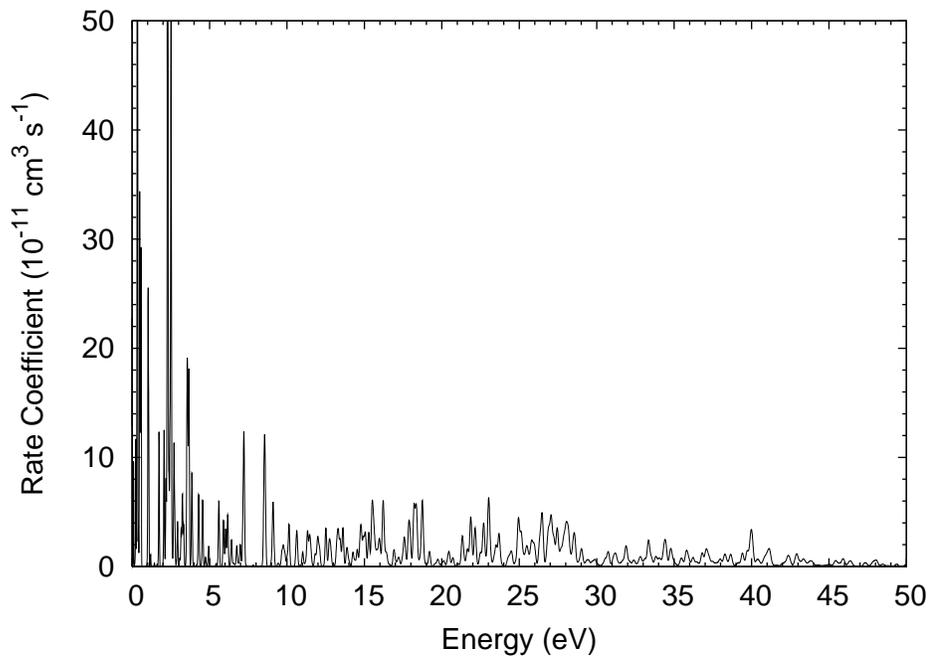}
\end{flushright}
\vspace{-.25in}
\caption{Velocity-convoluted DR cross sections for excitation
of the parent configuration 8. Solid curve, Basis B.
\label{fig-cf-8}} 
\end{figure}

In figures \ref{fig-cf-6} to \ref{fig-cf-8} we present results for core-excitations to parent
configurations 6--8, which are only accessible via configuration mixing. Again, there is
an allowed core-radiative stabilization pathway for limited $n$-values. We see that the
DR cross sections are progressively weaker as we go to higher-energy parent configurations.
The main difference now between the results of Basis A and Basis B, see figures \ref{fig-cf-6} and
\ref{fig-cf-7}, is an energy shift, due to the fact that we only adjusted to the lowest 40 observed
level energies.

We conclude that the use of observed energies mitigates against differences in the DR cross
section due to the different level energies of Basis A and Basis B while differences in
the radiative rates either occur for transitions which do not contribute strongly to the DR,
or the rates themselves are simple redistributed amongst near-by levels.

\subsection{Comparison with experiment}

We consider the $\Delta n=0$ and $\Delta n=1$ core-excitations separately. (Low-lying resonances
which arise from $n=3 - 4$ capture to $n=4$ do overlap the $\Delta n=0$ energy range but the
peaks are so small as to be `lost in the noise' when comparing with experiment.)

\subsubsection{$\Delta n=0$.}

\begin{figure}
\begin{flushright} 
\psfig{file=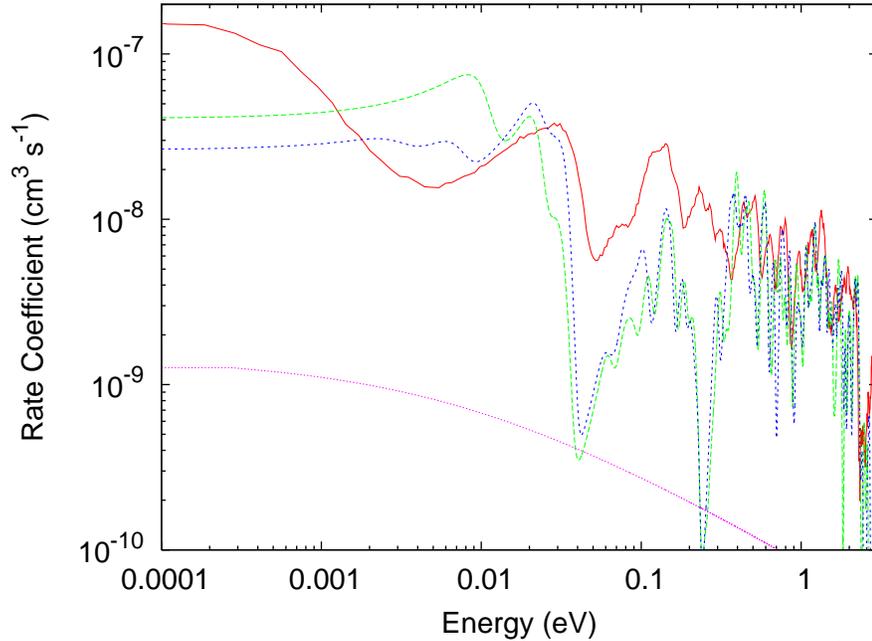}
\end{flushright}
\vspace{-.25in}
\caption{Velocity-convoluted recombination cross sections for Fe$^{13+}$ near threshold.
Solid (red) curve, experimental results of Schmidt \etal (2006); long-dashed (green) curve, 
theoretical DR results obtained on using Basis A; short-dashed (blue) curve, theoretical 
results obtained on using Basis B; dotted (purple) curve, theoretical RR results (Basis A). Theoretical 
results are all this work.
\label{fig-cf-9}} 
\end{figure}

In figure \ref{fig-cf-9} we compare our theoretical DR results, obtained using Basis A and Basis B,
with the experimental measurements of Schmidt \etal (2006). The comparison is made very close
to threshold ($10^{-4} - 1$~eV), utilizing a log-log scale. Schmidt \etal (2006) note a
relatively small anomalous enhancement below $10^{-3}$~eV and so estimate the `true' DR-plus-RR 
contribution to tend towards $1\times 10^{-7}$~cm${^3}$~s$^{-1}$ at $10^{-4}$~eV. The two sets of
theoretical results are in accordance above $\sim 0.01$~eV while at lower energies the
results of Basis B are in somewhat better agreement with experiment, down to $\sim 0.001$~eV.
Below the dip at about 0.04~eV, Schmidt \etal show fits to 5 DR resonances. We find 14
resonances, of the form $3\rms^2 3\rmp_{3/2} 32l (l=0,1) $, $3\rms 3\rmp^{2\, 4}P_{3/2} 9\rmf$, and
$3\rms3\rmp3\rmd 5\rmg$. Of course, our convoluted cross sections only exhibit 3 (Basis A) or 4
(Basis B) obvious peaks in this energy region.
We have looked at the bound states just below threshold but they are well below the apparent
uncertainties in resonance energy positioning seen in figure \ref{fig-cf-9}.
There is substantial disagreement between theory and experiment between about 0.03 and 0.35~eV.
Resonances in this region are higher members of those just series mentioned, as well as those
attached to parent configurations 6 and 7.
However, the differences below 0.2~eV have negligible effect on any differences in the
Maxwellian rate coefficient at 2~eV. Only the differences up to 0.35~eV start to impact upon the
Maxwellian rate coefficient at 2~eV. We also illustrate the RR contribution in this energy
region, having applied a hard cut-off at $n_\rmc=45$ again.

\begin{figure}
\begin{flushright} 
\psfig{file=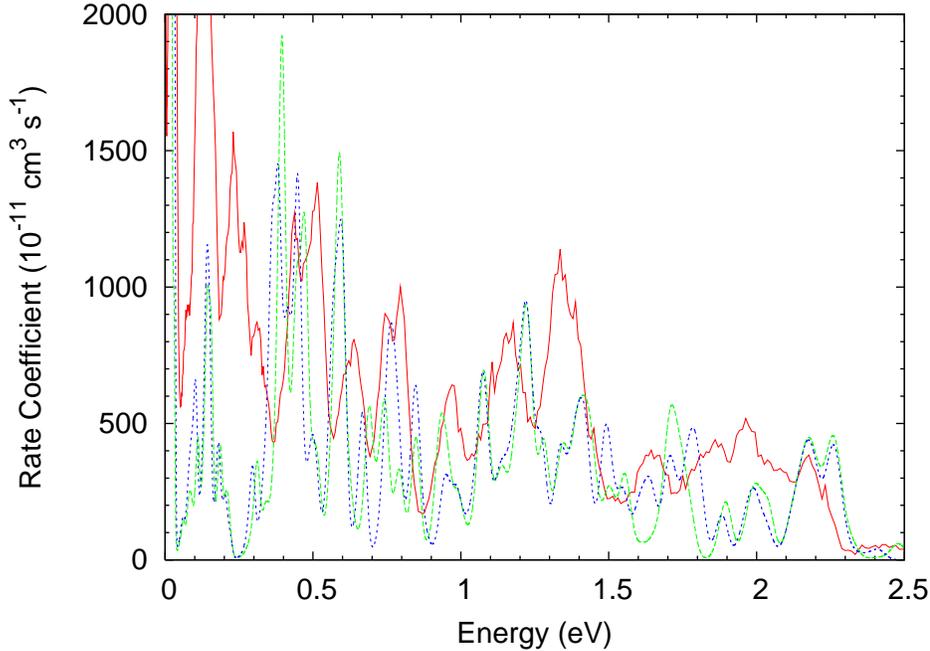}
\end{flushright}
\vspace{-.25in}
\caption{Velocity-convoluted $\Delta n=0$ recombination cross sections for Fe$^{13+}$ over $0-2.5$~eV.
Solid (red) curve, experimental results of Schmidt \etal (2006); long-dashed (green) curve, theoretical DR
results obtained on using Basis A; short-dashed (blue) curve, theoretical results obtained on using Basis B.
Theoretical results are all this work.
\label{fig-cf-10}} 
\end{figure}

In figure \ref{fig-cf-10} we compare the results of theory and experiment near threshold again,
this time using a linear plot. We see more clearly now that the theoretical results from Basis A and
Basis B are in close agreement over $\approx 0.05-0.4$~eV, but differ substantially from the
measured. Overall, in this energy range, the differences between the two sets of theoretical
results are not large enough to suggest an uncertainty which could account for the difference with the measured.
Although, where there are more noticeable differences, the results from Basis B are perhaps somewhat of an 
improvement over Basis A. The DR cross section drops substantially (by a factor of 10, or so)
above 2.34~eV because autoionization into the excited fine-structure level 2 becomes energetically 
allowed. This means that resonances below 2.34~eV contribute `disproportionately' at higher
temperatures --- see, e.g., figure 4 of Schmidt \etal (2006). If we sum-up the resonance strengths 
over 0.35 to 2.5~eV we find that the
result for Basis A is 3\% smaller than for Basis B but the measured is 33\% larger.

\begin{figure}
\begin{flushright} 
\psfig{file=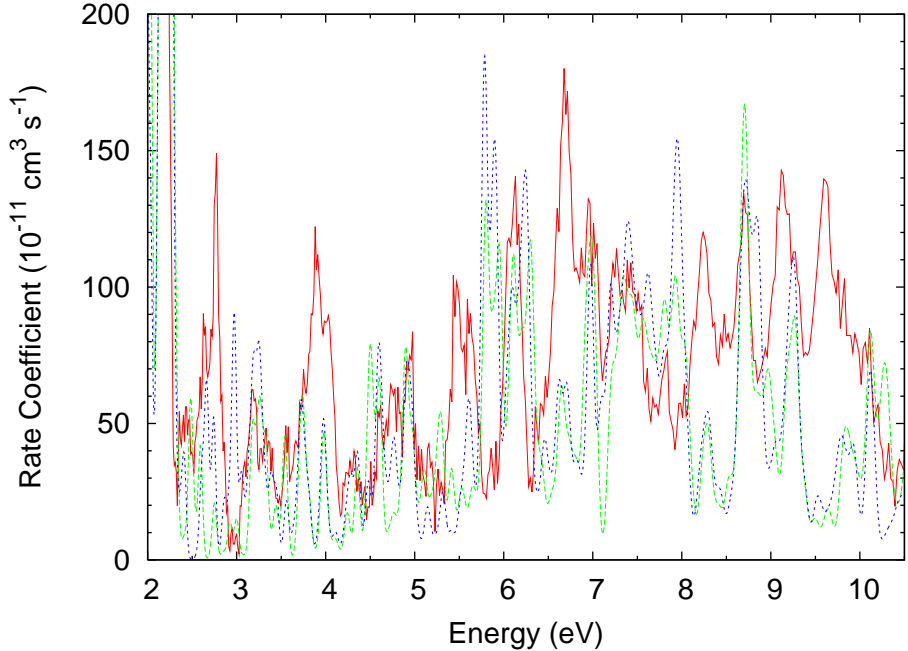}
\end{flushright}
\vspace{-.25in}
\caption{As figure 10, but over $2-10.5$~eV.
\label{fig-cf-11}} 
\end{figure}
In figure \ref{fig-cf-11} we compare the results of theory and experiment over $2-10$~eV.
Although this is the typical temperature range of peak abundance for Fe$^{13+}$ in a photoionized
plasma, only above a temperature corresponding to about 5~eV is the Maxwellian rate
coefficient dominated by the resonances above 2~eV. Again, the differences between the results
of Basis A and Basis B are not too significant, but the agreement with the measured is
rather poor. If we sum-up the resonance strengths over 2.5 to 10.5~eV we find that the
result for Basis A is 8\% smaller than for Basis B but the measured is 40\% larger.
Thus, over $2-10$~eV we expect the experimentally deduced Maxwellian rate coefficient to
be roughly 40\% larger than the theoretical one.

\begin{figure}
\begin{flushright} 
\psfig{file=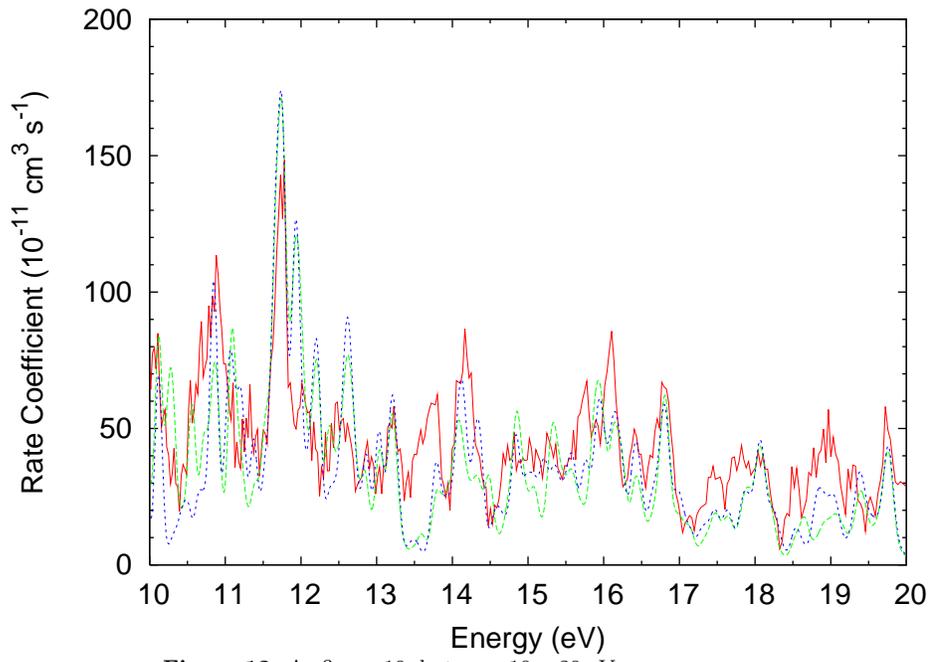}
\end{flushright}
\vspace{-.25in}
\caption{As figure \ref{fig-cf-10}, but over $10-20$~eV.
\label{fig-cf-12}} 
\end{figure}
\begin{figure}
\begin{flushright} 
\psfig{file=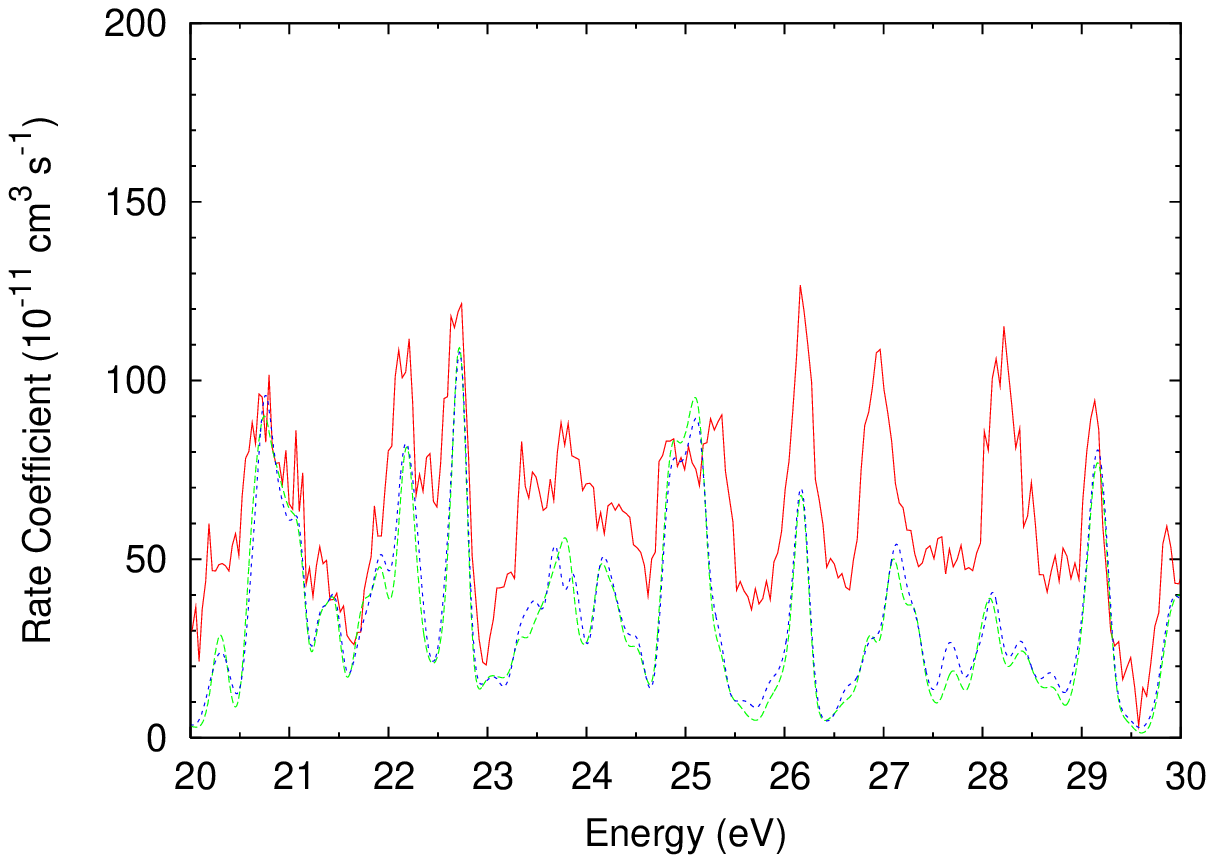}
\end{flushright}
\vspace{-.25in}
\caption{As figure \ref{fig-cf-10}, but over $20-30$~eV.
\label{fig-cf-13}} 
\end{figure}
\begin{figure}
\begin{flushright} 
\psfig{file=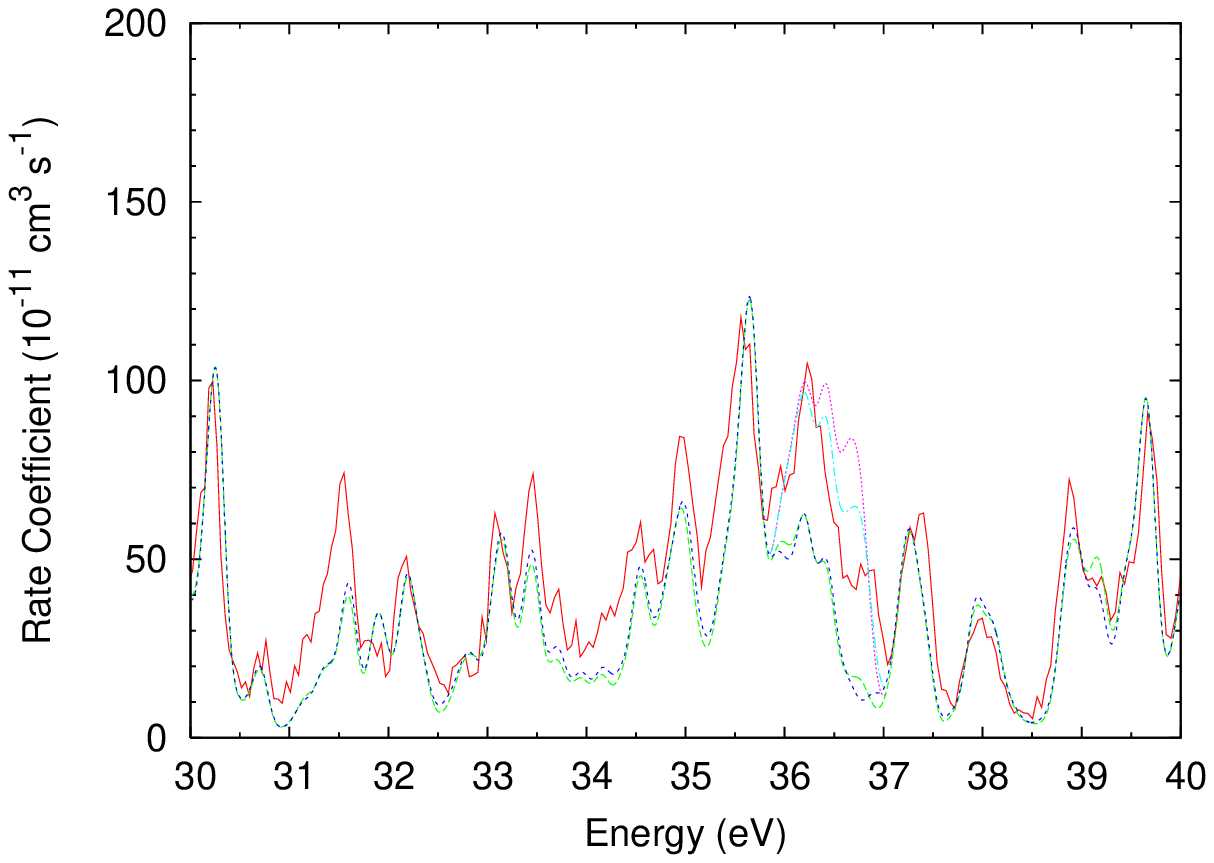}
\end{flushright}
\vspace{-.25in}
\caption{As figure \ref{fig-cf-10}, but over $30-40$~eV; plus: dotted (purple) curve, 
delayed cut-off time-of-flight 166.5~ns; dot-dashed (light blue) curve, 
utilizing the survival probabilities of Schippers \etal (2001).
\label{fig-cf-14}} 
\end{figure}

In figure \ref{fig-cf-12} we compare results over  $10-20$~eV. As we move up in energy we start to 
see more of a convergence between the theoretical results and the measured; the summed resonance 
strengths from Basis A and Basis B differ by less than 1\% whilst the measured is 21\% larger.
However, if we now look at the energy range $20-30$~eV (figure \ref{fig-cf-13}) we see that the
measured rate coefficient appears to be sitting on a much larger `background' over $23-29$~eV
--- the theoretical results drop much closer to zero between the main resonance peaks,
i.e., there is little `fill-in' due to other small resonance contributions. Over $30-40$~eV
(figure \ref{fig-cf-14}) the much better agreement between theory and experiment is resumed, 
except over $36-37$~eV. Here, there is a noticeable contribution from capture to high-$n$ states 
and the  comparison with experiment is dependent on (the modelling of) their survival to be detected. 
We illustrate with the results from two models (both for Basis A) and discuss
them in detail next.

\begin{figure}
\begin{flushright} 
\psfig{file=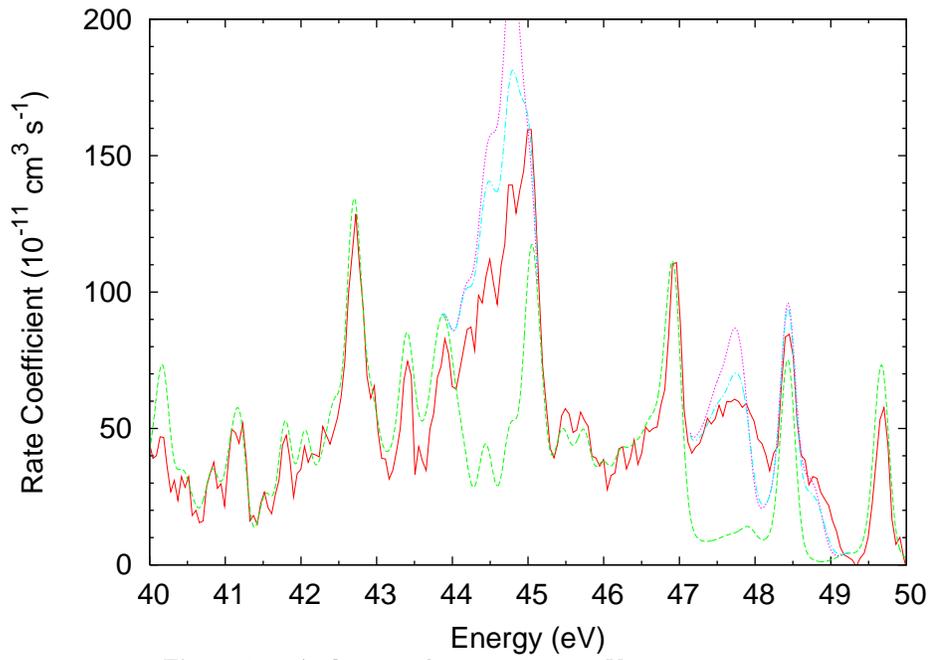}
\end{flushright}
\vspace{-.25in}
\caption{
As figure \ref{fig-cf-14}, but over $40-50$~eV.
\label{fig-cf-15}} 
\end{figure}
\begin{figure}
\begin{flushright} 
\psfig{file=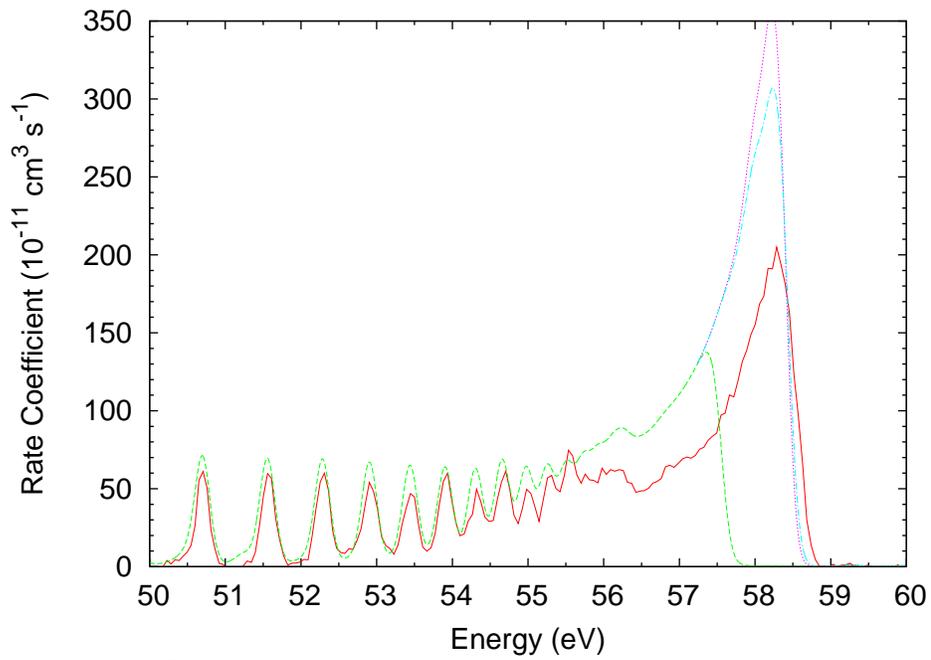}
\end{flushright}
\vspace{-.25in}
\caption{As figure \ref{fig-cf-14}, but over $50-60$~eV.
\label{fig-cf-16}} 
\end{figure}

In figures \ref{fig-cf-15} and \ref{fig-cf-16} we compare results over  $40-50$~eV and
$50-60$~eV, respectively. The resonances arise from $3\rms\rightarrow 3\rmp$ and
$3\rmp\rightarrow 3\rmd$ core-excitations for the former, but the second only for the
latter --- see figures \ref{fig-cf-2} and \ref{fig-cf-3}. The results of Basis A and
Basis B are indistinguishable on this scale, and so only Basis A results are shown and 
considered further. In these two energy regions we address the issue of the survival 
of the recombined states as the ions travel from the cooler to the charge-state-analyzer. 
Recall figure \ref{fig-cf-2},
there are high-$n$ DR contributions from parent levels $8-10$ of configuration 2 which span
$44-49$~eV. (Those attached to parent level 6 affect the $36-37$~eV range.) The situation, as
illustrated by figure \ref{fig-cf-3}, is simpler for the peak at 58~eV. It is in these two energy 
ranges for which a hard cut-off at $n_\rmc=45$ results-in large discrepancies between theory and
experiment. The discrepancy is reduced on implementing a delayed cut-off (equation \ref{prob}) 
utilizing the appropriate time-of-flight for this experiment of 166.5~ns and imposing a hard cut-off 
at $n_\rmc=95$ due to the correction magnet close to the cooler (Schippers, private communication).
Further improvement in agreement is obtained on utilizing the survival probabilities of Schippers 
\etal (2001), but for the DR of Fe$^{13+}$. These latter two sets of results are only shown at 
energies where they differ from the `hard cut-off' results. Clearly, the final agreement between  
theory and experiment is sensitive to the precise contribution, i.e. survival, of recombined states 
with $n>45$. While this largely accounts for the discrepancies over $44-49$~eV between experiment and 
theory utilizing only a hard cut-off, as well as the mis-match in the final position of the 
Rydberg peak at 58.5~eV, it has no effect on the puzzling discrepancy between about 56 and 58eV. 
Here, the experimental result actually lies at a fairly uniform $4\times 10^{-10}$~cm$^3$s$^{-1}$ 
below (all of) the theoretical one(s).

We close the discussion of the $\Delta n=0$ results with a small observation: the noticeable drop 
in the DR cross section just below 56.5~eV  (in figure \ref{fig-cf-16}) is due to the final-state of the 
$3\rmd_{3/2}\rightarrow 3\rmp_{3/2}$ core-radiative stabilization pathway infact opening-up at this point 
($n=32$) to autoionization to the $3\rmp_{1/2}$ continuum. (We obtain 0.220 for the ratio of the
$3\rmd_{3/2}\rightarrow 3\rmp_{3/2}$ to $3\rmd_{3/2}\rightarrow 3\rmp_{1/2}$ radiative rates compared 
to 0.225 obtained by Storey \etal (2000) --- see also table \ref{tab2}.)
Thus, there appears to be close qualitative agreement between theory and experiment
for this effect. This is in stark contrast to the poor quantitative agreement for the absolute cross section.

\subsubsection{$\Delta n=1$.}
\begin{figure}
\begin{flushright} 
\psfig{file=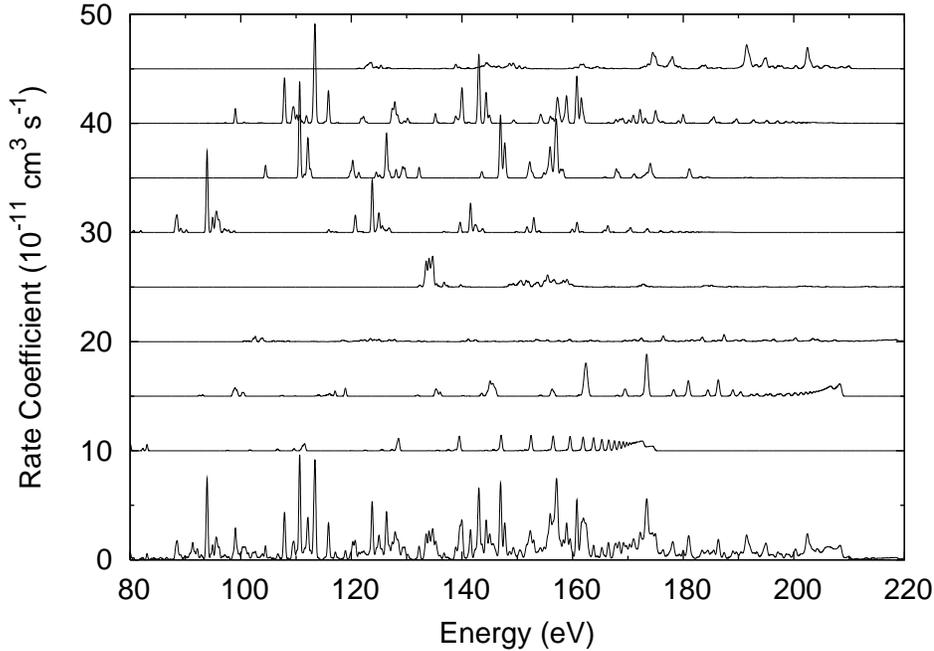}
\end{flushright}
\vspace{-.25in}
\caption{Velocity-convoluted DR cross sections for $3-4$ core-excitations in Fe$^{13+}$.
Bottom curve, summed-over all parent configurations. Offset (increasing) are the
contributions from $3\rmp \rightarrow 4\rms, 4\rmd$; $3\rms \rightarrow 4\rmp, 4\rmf$; 
$3\rmp \rightarrow 4\rmp, 4\rmf$; and $3\rms \rightarrow 4\rms, 4\rmd$ core-excitations.
\label{fig-cf-17}} 
\end{figure}

In figure \ref{fig-cf-17} we present an overview of the different contributions to the 
complete $3-4$ `spectrum'.
The first point to note is that the resonance strengths are now a factor 10,
or more, smaller than those we have seen associated with the $\Delta n=0$ core-excitations.
Only the $3\rmp \rightarrow 4\rms, 4\rmd$ core-excitations exhibit the classic DR spectrum.
The dipole core-excitations dominate, along with $3\rms \rightarrow 4\rmd$.
We see that the final total spectrum is complex, which makes it difficult to identify
individual peaks in the measured spectrum.

\begin{figure}
\begin{flushright} 
\psfig{file=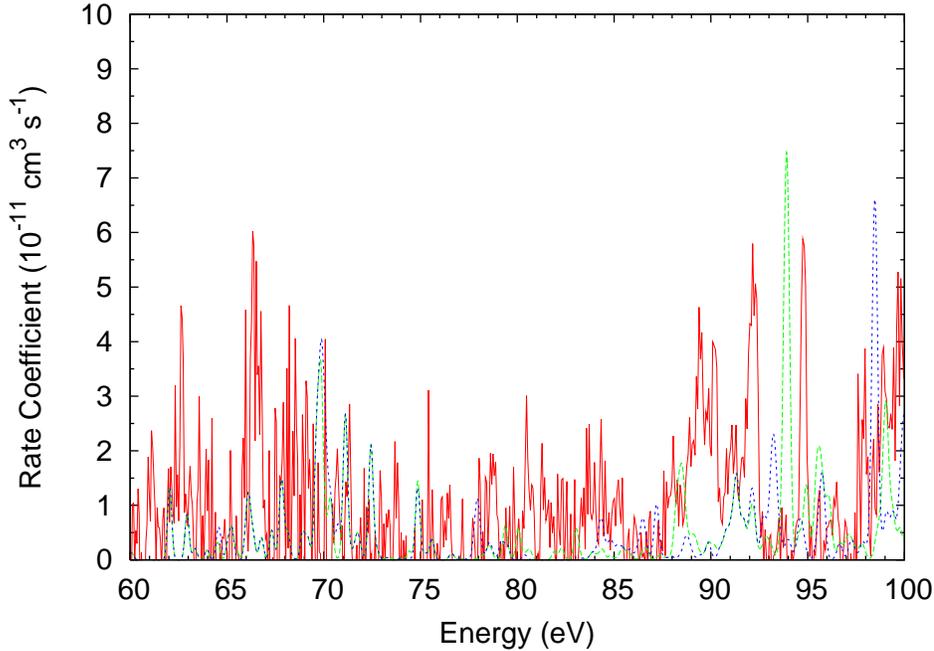}
\end{flushright}
\vspace{-.25in}
\caption{Velocity-convoluted $\Delta n=1$ $(3-4)$ DR cross sections for Fe$^{13+}$
over $60-100$~eV. 
Solid (red) curve, experimental results of Schmidt \etal (2006); long-dashed (green) curve, theoretical DR
results obtained on using from Basis C; short-dashed (blue) curve; theoretical results from Basis D.
Theoretical results are all this work.
\label{fig-cf-18}} 
\end{figure}

\begin{figure}
\begin{flushright} 
\psfig{file=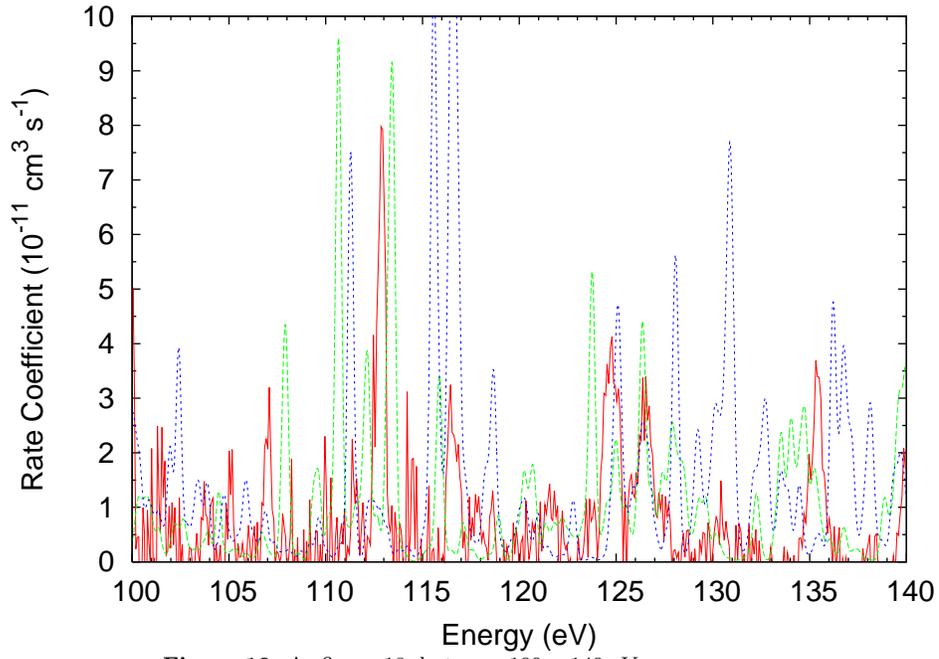}
\end{flushright}
\vspace{-.25in}
\caption{As figure \ref{fig-cf-18}, but over $100-140$~eV.
\label{fig-cf-19}} 
\end{figure}

\begin{figure}
\begin{flushright} 
\psfig{file=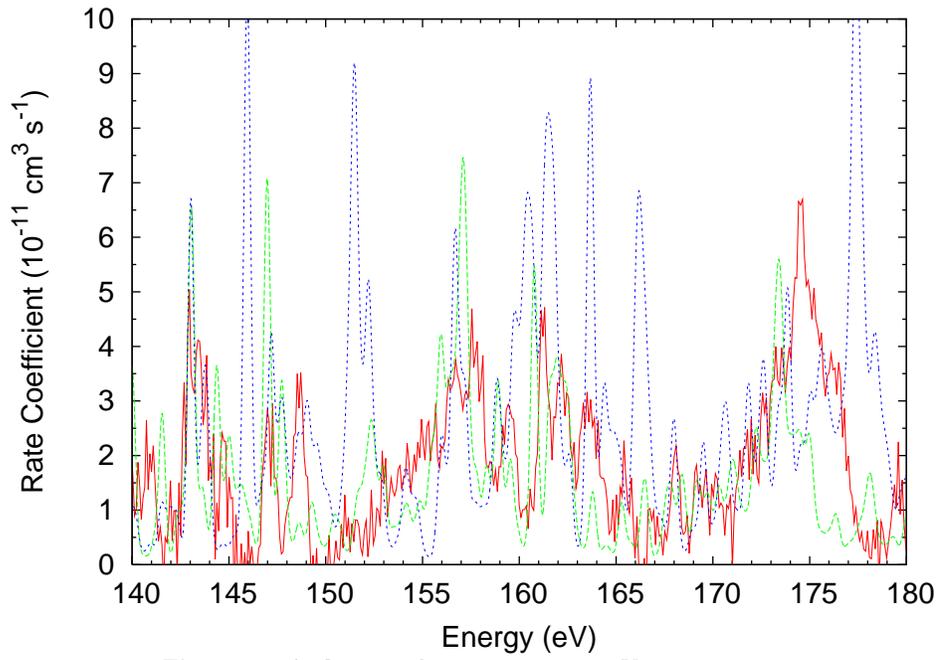}
\end{flushright}
\vspace{-.25in}
\caption{As figure \ref{fig-cf-18}, but over $140-180$~eV.
\label{fig-cf-20}} 
\end{figure}

\begin{figure}
\begin{flushright} 
\psfig{file=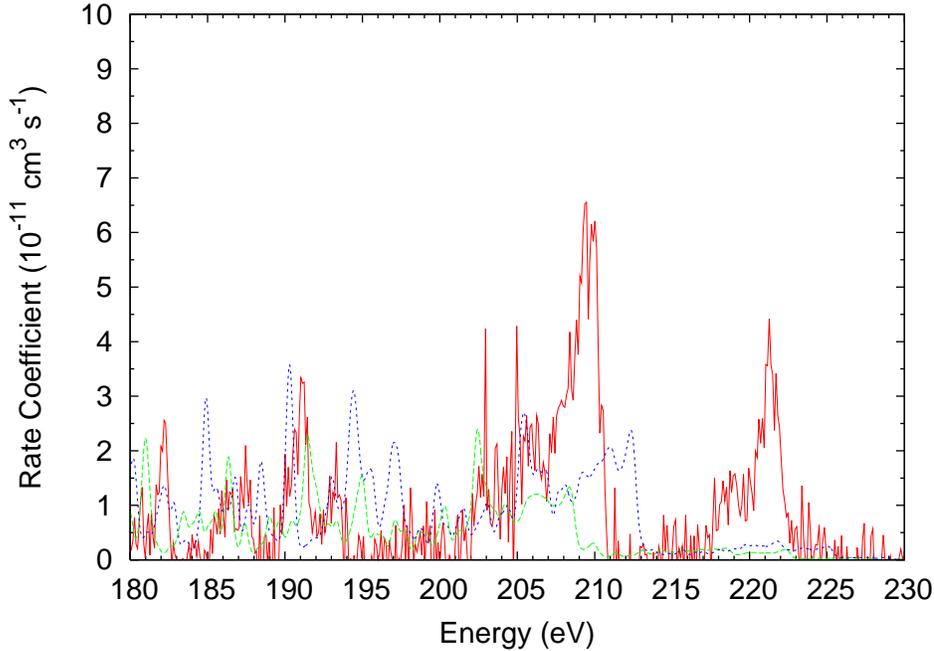}
\end{flushright}
\vspace{-.25in}
\caption{As figure \ref{fig-cf-18}, but over $180-230$~eV.
\label{fig-cf-21}} 
\end{figure}

In figures \ref{fig-cf-18}--\ref{fig-cf-21} we make comparisons between the results of our calculations with those of the
experimental measurements by Schmidt \etal (2006) for DR in the energy region in which the $3-4$ core-excitations
contribute. 

In figure \ref{fig-cf-18} it is capture to $n=4$ which dominates, and we did not make a separate
calculation using Basis D since the orbitals were based on the Thomas--Fermi potential
optimized for the Al-like core. In the case of Basis C, although we used the same scaling parameters 
for capture to $n=4$ as for $n>4$, the  Slater-Type-Orbital model potential depends on configuration 
(actually, the complex) and so is inherently `optimized' differently for capture to $n=4$ and $n>4$
(see Burgess \etal 1989 for specific details).
Hence, we added the Basis C $n=4$ results to those of Basis D. In the $60-90$~eV range the theoretical 
cross sections are much weaker than the measured. 

In contrast, in figures \ref{fig-cf-19} and \ref{fig-cf-20} we see that the theoretical DR cross sections 
are more strongly peaked than the measured, especially so for Basis D which has the larger 
$4\rightarrow 3$ radiative rates. Despite Basis C apparently giving rise to a worse structure 
for Fe$^{13+}$ than Basis D, based upon the agreement between the length and velocity forms
for the relevant oscillator strengths (see table \ref{tab4}), it does appear that Basis C gives rise to 
distinctly better agreement with experiment for the DR cross sections than Basis D. Although the agreement
is worse than for $\Delta n=0$ core-excitations, the sensitivity of the atomic structure to the overlap
of the $n=3$ and $n=4$ orbitals means that the differences are `less serious', i.e., there is still
enough uncertainty in the atomic structure so that the agreement might be improved upon. We note that
there is no simple identification of the various peaks in these figures as many different core-excitations 
contribute in this energy region, but they are dominated by a few low-$n$ resonances because $\Delta n=0$ 
autoionization of the final states opens-up at $n<10$ for all parents, except $3\rms^2 3\rmp$. In addition, 
the $3\rmp \rightarrow 4\rms$ series Rydberg accumulation contributes to the peak at 175~eV.

In figure \ref{fig-cf-21} we can unambiguously identify the peaks around 210 and 220~eV as being associated
with the limit of the $3\rmp\rightarrow 4\rmd$  and $3\rms \rightarrow 4\rmp$ core-excitations, but the theoretical
cross sections are much smaller than measured for these peaks, especially so for the latter.
The sum over $n$ is fairly well converged by $n=45$ and even applying no cut-off (or infinite time-of-flight)
does not increase the size of the theoretical peaks by much. We do see the `overshoot' of the
Basis D results here --- as expected from the use of the unadjusted energies of Table \ref{tab3}.

Summing over all DR resonance strengths for the $3-4$ core-excitations we find the Basis D result
to be 50\% larger than that for Basis C, which is comparable with the excess of the $3\rmp - (4\rms + 4 \rmd)$ 
oscillator strength seen in Table \ref{tab4}. The sum of the measured DR resonance strengths in the
$60-240$~eV range is 21\% larger than that from Basis C. Of course, we have noted significant disagreements
between the calculated and measured DR resonances, both over- and under-estimates.

\begin{figure}
\begin{flushright} 
\psfig{file=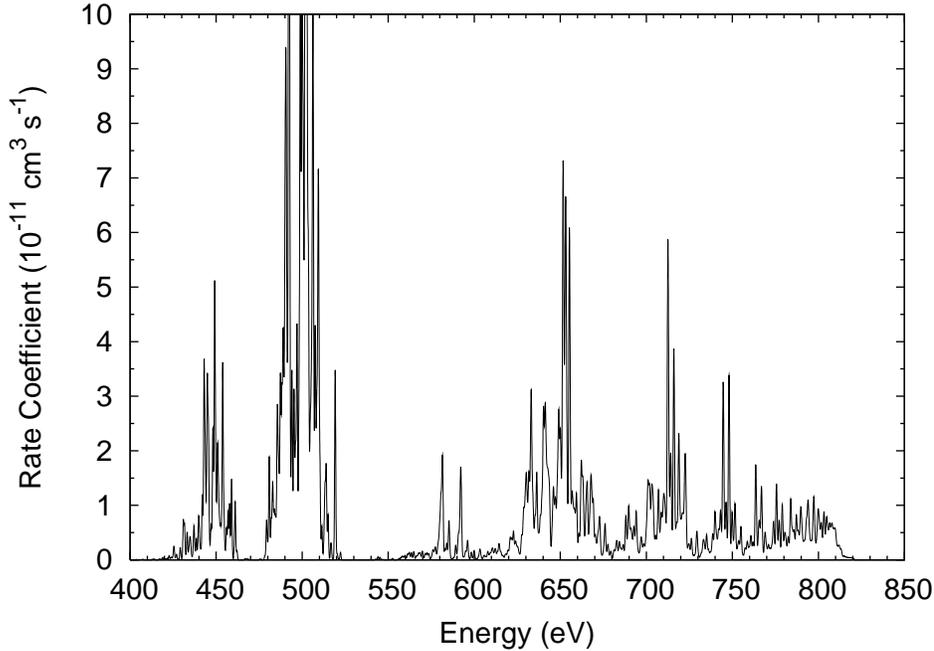}
\end{flushright}
\vspace{-.25in}
\caption{Velocity-convoluted $\Delta n=1$ $(2-3)$ recombination cross sections for Fe$^{13+}$.
Solid curve, present theoretical results.
\label{fig-cf-22}} 
\end{figure}

Finally, in figure \ref{fig-cf-22} we present our results for the $2-3$ core-excitations. These resonances
lie above the highest energy considered by Schmidt \etal (2006). It should be noted, however, that the sum of 
DR resonance strengths associated with this $2-3$ core-excitation is a factor of 2.5 larger than that associated
with the $3-4$. Thus, apart from contributing at a higher temperature, this core-excitation is likely
to be more important than the $3-4$ for application to collisional plasmas.

\section{Maxwellian rate coefficients}
\begin{figure}
\begin{flushright} 
\psfig{file=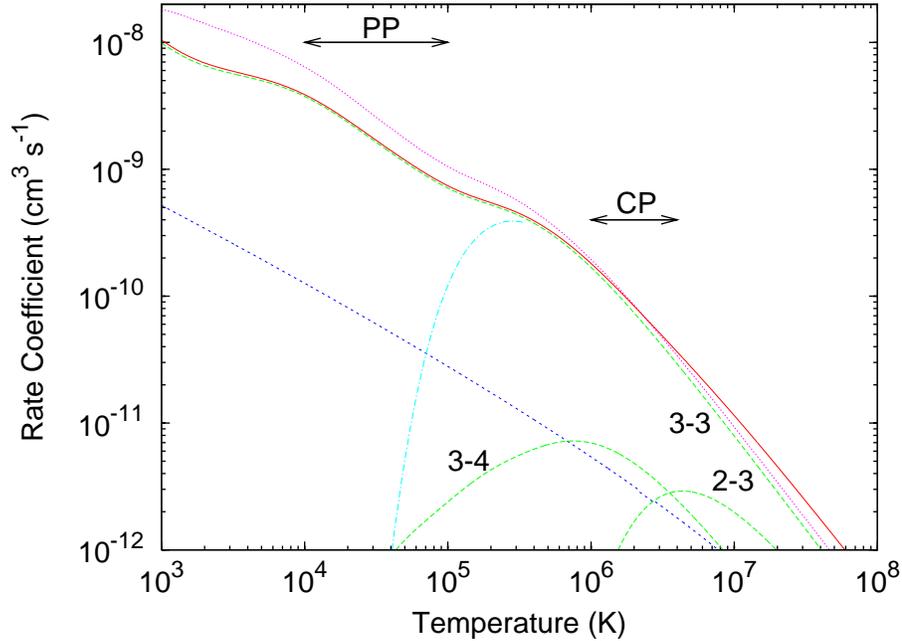}
\end{flushright}
\vspace{-.25in}
\caption{Maxwellian rate coefficients for Fe$^{13+}$. Solid (red) curve, total DR-plus-RR; short-dashed (blue) curve, RR;
long-dashed (green) curves, DR for $3-3$, $3-4$ and $2-3$ core-excitations. All this work.
Dot-dashed (light blue) curve, DR of Arnaud and Raymond (1992). Dotted (purple) curve,
experimentally based total of Schmidt \etal (2006). PP and CP denote typical photoionized and electron collisional
plasma temperature ranges, respectively, for Fe$^{13+}$ (Kallman and Bautista 2001 and Mazzotta \etal 1998). 
\label{fig-cf-23}} 
\end{figure}

In figure \ref{fig-cf-23} we present our theoretical results for Maxwellian rate coefficients: RR-plus-DR from
$3-3$, $3-4$ and $2-3$ core-excitations, and compare the sum total of these with the one determined by
 Schmidt \etal (2006), based primarily upon their measured DR (cooler) rate coefficients. Over $10^4-10^5$~K,
a typical temperature range for photoionized plasmas where Fe$^{13+}$ is abundant, the experimentally based total 
is between a factor $1.52 - 1.38$ larger than our theoretical total --- this is inline with what we expect following 
our earlier detailed comparison of the DR resonances contributing at these temperatures. Nevertheless, it is
clear that the total recombination rate coefficient of Fe$^{13+}$ in photoionized plasmas is an order
of magnitude larger than has been used to-date, as first pointed-out by Schmidt \etal (2006) on the
basis of their measurements for this ion. We show also only the low temperature fall-off of the recommended 
DR rate coefficient of Arnaud and  Raymond (1992) and not any of the {\it ad hoc} changes proposed by 
Netzer (2004) and by Kraemer \etal (2004).

At temperatures of a few times $10^6$~K, typical of electron collision dominated plasmas where Fe$^{13+}$ is abundant, 
the  experimentally based total  is only 5\% smaller than our calculated one, while the recommended data of Arnaud 
and Raymond (1992) lies  only about 10\% higher. (The data of Arnaud and Raymond (1992) is based principally upon the 
results of Jacobs \etal (1977), but includes an estimate of the contribution from $2\rmp - 3\rmd$ inner-shell
transitions as well and which were not included by Jacobs \etal.) We see also that both $\Delta n=1$ contributions 
contribute only modestly
to the total, both equally about 6\% at $3\times 10^6$~K. Far off equilibrium, this rises to about 20\% at $10^7$~K
with three-quarters coming from the $2-3$ core-excitation. Given the modest contribution from $\Delta n=1$
core-excitations, then, since such calculations are more demanding than for $\Delta n=0$, the use of
$LS$-coupling may suffice, even with the 30\% difference from intermediate coupling which we note (not shown).

\begin{figure}
\begin{flushright} 
\psfig{file=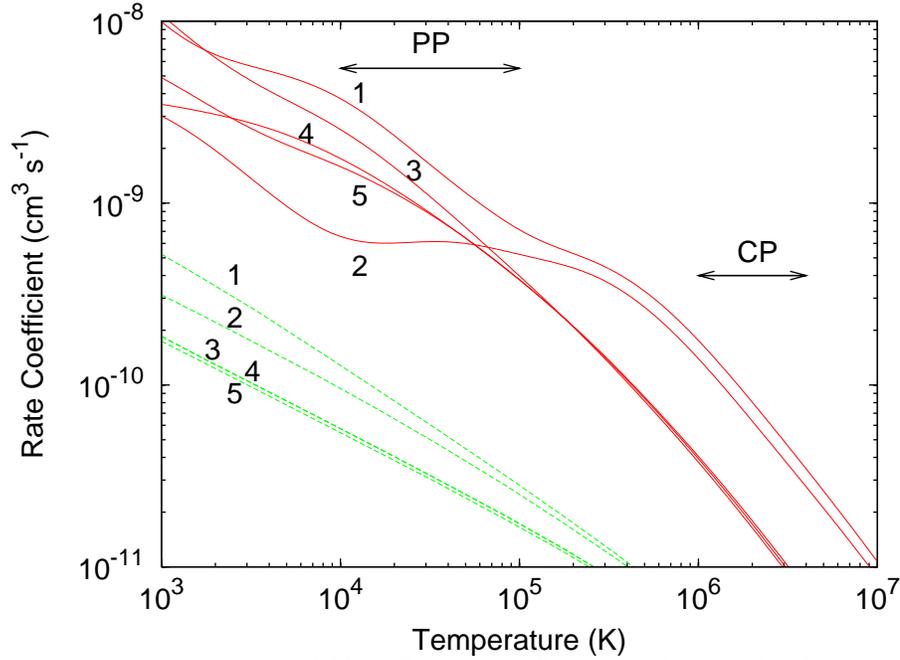}
\end{flushright}
\vspace{-.25in}
\caption{Total Maxwellian rate coefficients for the ground and metastable initial levels ($m=1-5$)
of Fe$^{13+}$ (see text for details). Solid (red) curves, DR; dashed (green) curves, RR. 
All this work. 
PP and CP, as figure \ref{fig-cf-23}.
\label{fig-cf-24}} 
\end{figure}

These rate coefficients are for the $3\rms^2 3\rmp ^2\rmP_{J=1/2}$ ground level of Fe$^{13+}$.
In photoionized plasmas, the ion population may not be concentrated in the ground level, while in electron 
collisional plasmas there may be signification population of levels of the $3\rms 3\rmp^{2\,\,4}\rmP$ term.
The rate coefficients for the various metastable levels can be quite different. In figure \ref{fig-cf-24}
we compare and contrast total DR (i.e., summed-over all core-excitations) and RR rate coefficients from
the ground and metastable levels. We index the target metastable  levels by $m$, where $m=1$ for the 
($J=1/2$) ground level, $m=2$ for the other ($J=3/2$) fine-structure level and $m=3, 4, 5$ for the 
($J=1/2, 3/2, 5/2$) levels of the metastable term, as per table \ref{tab1}.

At photoionized plasma temperatures we see that the $m=2$ metastable DR rate coefficient is an
order of magnitude smaller than for the ($m=1$) ground level --- this is due primarily, of course, to the absence
of the fine structure DR pathway. The DR rate coefficients for higher metastables exhibit irregular
behaviour (at low temperatures) due to the positioning close to threshold of the lowest autoionizing states,
relative to these excited Fe$^{13+}$ thresholds, although the $m=3$ and 4 metastable levels again
have fine-structure dielectronic capture pathways and the enhancement for $m=3$ appears to be quite
noticeable. At collisional plasma temperatures the DR rate
coefficients split primarily into two groups which are based upon term, not level, as the influence of
fine-structure DR and threshold effects is diminished.

The difference in RR rate coefficients is much less pronounced, at all temperatures. At low temperatures
the stability against autoionization for all $n$ distinguishes RR of the ground level. There is little
difference for metastables levels of the excited term because the highest stable recombined $n$ is
(almost) independent of the fine-structure parent.

\begin{table}
\caption{DR fitting coefficients $c_i$~(cm$^3$s$^{-1}$K$^{3/2}$) and $E_i$(K) for the ground and 
metastable levels  ($m=1-5$) of Fe$^{13+}$.\label{tab5}}
\footnotesize
\begin{tabular}{ccccccccc}
\br
$m$&$c_1$&$c_2$&$c_3$&$c_4$&$c_5$&$c_6$&$c_7$&$c_8$\\
\mr
1&1.090($-$3)&7.801($-$3)&1.132($-$2)&4.740($-$2)&1.990($-$1)&3.379($-$2)&1.140($-$1)&1.250($-$1)\\
2&3.176($-$4)&1.097($-$3)&1.451($-$2)&4.623($-$2)&1.424($-$1)&3.105($-$2)&1.173($-$1)&1.579($-$1)\\
3&9.230($-$4)&4.787($-$3)&7.598($-$3)&1.538($-$2)&1.512($-$2)&1.711($-$2)&9.083($-$3)&4.875($-$1)\\
4&6.837($-$4)&3.386($-$3)&8.737($-$3)&2.334($-$2)&2.819($-$2)&1.282($-$2)&9.735($-$3)&2.670($-$1)\\
5&5.606($-$4)&3.306($-$3)&9.372($-$3)&1.635($-$2)&1.674($-$2)&1.783($-$2)&9.195($-$3)&3.689($-$1)\\
\mr

$m$&$E_1$&$E_2$&$E_3$&$E_4$&$E_5$&$E_6$&$E_7$&$E_8$\\
\mr
1&1.246(3)&1.063(4)&4.719(4)&1.952(5)&5.637(5)&2.248(6)&7.202(6)&3.999(9)\\
2&1.204(3)&1.214(4)&5.689(4)&1.983(5)&5.340(5)&2.414(6)&7.302(6)&4.245(9)\\
3&9.887(2)&1.075(4)&4.542(4)&1.832(5)&5.506(5)&1.692(6)&6.994(6)&2.789(9)\\
4&1.824(3)&1.101(4)&4.775(4)&2.131(5)&1.109(6)&5.479(6)&1.365(9)&2.517(9)\\
5&1.288(3)&1.171(4)&4.780(4)&1.828(5)&5.530(5)&1.697(6)&6.960(6)&2.666(9)\\
\br
\end{tabular}
\end{table}
\normalsize

\begin{table}
\caption{RR fitting coefficients for the ground and metastable levels ($m=1-5$) of Fe$^{13+}$.\label{tab6}}
\begin{indented}
\item[]\begin{tabular}{rrrrrrr}
\br
 $m$ & $A$~(cm$^3$s$^{-1}$) &  $B$  &$T_0$(K) & $T_1$(K) & $C$ & $T_2$(K)\\
\mr
  1 &  4.321($-$10) &  0.6091 &  2.255(03) &  4.962(07) &  0.0356 &  1.006(05) \\
  2 &  2.031($-$11) &  0.5464 &  2.669(05) &  5.310(07) &  0.0277 &  9.907(08) \\
  3 &  1.591($-$09) &  1.0274 &  1.196(01) &  2.038(07) &  0.0449 &  1.764(08) \\
  4 &  1.591($-$09) &  1.0274 &  1.196(01) &  2.038(07) &  0.0449 &  1.764(08) \\
  5 &  4.803($-$11) &  0.3781 &  1.920(04) &  4.025(07) &  0.5220 &  5.410(03) \\
\br
\end{tabular}
\end{indented}
\end{table}

In tables \ref{tab5} and \ref{tab6} we present separately the fitting coefficients for our total DR and RR
rate coefficients for the ground and metastable levels (indexed by $m$), which are based upon the functional 
forms given by equations (\ref{DRfit}) and (\ref{RRfit}), respectively. The fits are accurate to better
than 1\% over $z^2 (10^1 - 10^7)$ K, where $z=13$ here.

\section{Summary}
\label{sum}
We have carried-out a series of multi-configuration Breit--Pauli calculations for the
dielectronic recombination of Fe$^{13+}$. Whilst there is much agreement between the
theoretical velocity-convoluted cross sections and those determined experimentally by 
Schmidt \etal (2006), differences 
over $0.1-10$~eV lead to the experimentally based total Maxwellian recombination rate coefficient
being upwards of 50\% larger than the theoretical one over the temperature range $10^4-10^5$~K,
which is typical of photoionized plasmas where Fe$^{13+}$ is abundant. Such a difference lies 
well outside of the theoretical uncertainty, based-upon the accuracy of the radiative rates
and sensitivity to resonance positions. It is also well outside of the experimental
uncertainty of $\pm 18\%$ (Schmidt \etal 2006). 

It is difficult to see how to resolve this difference. Simply carrying-out a larger (configuration
interaction) calculation would not be expected to result in a change much beyond the difference, 
already  noted, between the Basis A and Basis B results, especially given the level of
agreement for radiative rates which we have observed between Bases A and B and the
extended Basis 2 of Storey \etal (1996). Furthermore, the good agreement between 
theory and experiment at higher $\Delta n=0$ energies is interrupted twice by disconcerting differences
spanning several electron volts. Such differences are also outside of the range of higher-order
effects such as interacting resonances and the interference between DR and RR.
Perhaps the results of a separate, independent, calculation will shed some light on the matter. 

Nevertheless, the theoretical recombination rate coefficient determined here for Fe$^{13+}$ is an 
order of magnitude larger than has been used by modellers in the past.
This may help explain the discrepancy between the iron M-shell ionization balance 
predicted by photoionization modelling codes 
and that 
deduced from the iron M-shell
unresolved-transition-array absorption feature observed in the X-ray spectra of many active galactic nuclei.
New data are clearly required for the other
Fe $3\rmp^q$ ions, especially $q=2-5$, in order to eliminate the uncertainty in the DR atomic
data used by {\sc cloudy}, {\sc ion} and {\sc xstar} and to enable them to focus on the `bigger picture'. 

\section*{Acknowledgments}
I would like to thank Eike Schmidt for supplying the experimental results in
numerical form, Stefan Schippers for providing the survival probabilities applicable
to the experimental set-up and Daniel Savin for commenting on an early draft of the manuscript. 
This work was supported in part by PPARC Grant No.
PPA$\backslash$G$\backslash$S2003$\backslash$00055 with the University of Strathclyde.

\References

\item[] Arnaud M and Raymond J 1992 \APJ {\bf 398} 394--406
\item[] Badnell N R 1987 \JPB {\bf 19} 3827--35
\item[] Badnell N R 1997 \jpb {\bf 30} 1--11
\item[] Badnell N R 2006 \APJS Submitted, astro-ph/0604144
\item[] Badnell N R, O'Mullane M G, Summers H P, Altun Z, Bautista M A, Colgan J, Gorczyca T W,
        Mitnik D M, Pindzola M S and Zatsarinny O 2003 {\it Astron. Astrophys.} {\bf 409} 1151--65
\item[] Badnell N R and Seaton M J 2003 \jpb {\bf 36} 4367--85
\item[] Bethe H A and Salpeter E E 1957 {\it Quantum Mechanics of One- and Two-Electron Atoms} (Springer-Verlag: Berlin)
\item[] Bryans P, Badnell N R, Gorczyca T W, Laming J M, Mitthumsiri W and Savin D W 2006 \APJS At Press, astro-ph/0604363
\item[] Burgess A 1964 \APJ {\bf 139} 776--80
\item[] Burgess A, Mason H E and Tully J A 1989 \AA {\bf 217} 319--28
\item[] Damburg R J and Kolosov V V 1979 \JPB {\bf 12} 2637--43
\item[] Dittner P F, Datz S, Miller P D and Pepmiller P L 1986 \PR A {\bf 33} 124--30
\item[] Ferland G J Private communication/To be submitted.
\item[] Fogle M, Badnell N R, Ekl\"{o}w N, Mohamed T. and Schuch R 2003 \AA {\bf 409} 781--6
\item[] Fogle M, Badnell N R, Glans P, Loch S D, Madzunkov S, Abdel-Naby Sh A,
Pindzola M S and Schuch R 2005 \AA {\bf 442} 757--66
\item[] Gorczyca T W, Badnell N R and Savin D W 2002 \PR A {\bf 65} 062707(8)
\item[] Gu M F 2003 \APJ {\bf 590} 1131--40
\item[] Jacobs V L, Davis J, Kepple P C and Blaha M 1977 \APJ {\bf 211} 605--11
\item[] Kallman T and Bautista M B 2001 \APJS {\bf 133} 221--53
\item[] Kraemer S B, Ferland G J and Gabel J R 2004 \APJ {\bf 604} 556--61
\item[] Linkemann J, Kenntner J, M\"{u}ller A, Wolf A, Habs D, Schwalm D, Spies W, Uwira O,
Frank A, Liedtke A, Hofmann G, Salzborn E, Badnell N R and Pindzola M S 1995 
{\it Nucl. Instrum. Methods Phys. Res.} B {\bf 98} 154--7
\item[] Mazzotta P, Mazzitelli G, Colafrancesco S and Vittorio N 1998 \AASS {\bf 133} 403--9
\item[] M\"{u}ller A 1999 {\it Int. J. Mass Spectrom.} {\bf 192} 9--22
\item[] Netzer H 2004 \APJ {\bf 604} 551--5
\item[] Netzer H, Kaspi S, Behar E, Brandt W N, Chelouche D, George I M, Crenshaw D M,
Gabel J R, Hamann F W, Kraemer S B, Kriss G A, Nandra K, Peterson B M, Shields J C and Turner T J
2003 \APJ {\bf 599} 933--48
\item[] NIST Atomic Spectra Database v3.0.3 2006 http://physics.nist.gov/PhysRefData/ASD
\item[] Pindzola M S, Badnell N R and Griffin D C 1992 \PR A {\bf 46} 5725--29
\item[] Savin D W, Bartsch T, Chen M H, Kahn S M, Liedahl D A, Linkemann J, M\"{u}ller A,
Schippers S, Schmitt M, Schwalm D and Wolf A 1997 \APJ {\bf 489} L115--8
\item[] Savin D W, Gwinner G, Grieser M, Repnow R, Schnell M, Schwalm D, Wolf A, Zhou S-G,
Kieslich S, M\"{u}ller A, Schippers S, Colgan J, Loch S D, Badnell N R, Chen M H and Gu M F
2006 \APJ {\bf 642} 1275--85
\item[] Schippers S, M\"{u}ller A, Gwinner G, Linkemann J, Saghiri A A and Wolf A 2001
\APJ {\bf 555} 1027--37.
\item[] Schmidt E W, Schippers S, M\"{u}ller A, Lestinsky M, Sprenger F, Grieser M, Repnow P,
Wolf A, Brandau C, Luki\'{c} D, Schnell M and Savin D W 2006 \APJ {\bf 641} L157--60
\item[] Storey P J, Mason H E and Saraph H E 1996 \AA {\bf 309} 677--82
\item[] Storey P J, Mason H E and Young P R 2000 \AASS {\bf 141} 285--96
\item[] Verner D A and Ferland G J 1999 \APJS {\bf 103} 467--73
\item[] Zong W, Schuch R, Gao H, DeWitt D R and Badnell N R 1998 \jpb {\bf 31} 3729--42
\endrefs

\end{document}